\documentclass[11pt]{scrartcl}

\KOMAoptions{bibliography=totoc}  %

\usepackage[dvipsnames]{xcolor}
\usepackage{graphicx}
\usepackage{amsmath}
\usepackage{amssymb}
\usepackage{units}
\usepackage{cancel}
\usepackage{hyperref}
\usepackage{todonotes}
\usepackage{booktabs}
\usepackage{tabularx}
\usepackage{upgreek}
\usepackage{ifthen}
\usepackage[square,numbers,sort&compress]{natbib}

\presetkeys{todonotes}{inline}{}
\numberwithin{equation}{section}

\hypersetup{
    colorlinks=true,
    citecolor={OliveGreen},
    linkcolor={Maroon},
}

\newcommand{\muline}[1]{\underline{\smash{#1}\vphantom{T}}\vphantom{#1}}

\newcommand{\vect}{\boldsymbol}
\newcommand{\abs}[1]{|#1|}
\newcommand{\mean}[1]{\langle #1\rangle}
\newcommand{\diff}{\mathrm{d}}

\newcommand{\kBT}{k_\mathrm{B} T}
\newcommand{\Vsys}{V_\mathrm{sys}}
\newcommand{\Rcrit}{R_\mathrm{c}}

\newcommand{\Eqref}[1]{Eq.~\eqref{#1}}
\newcommand{\Eqsref}[1]{Eqs.~\eqref{#1}}
\newcommand{\figref}[1]{Fig.~\ref{#1}}
\newcommand{\secref}[1]{section~\ref{#1}}
\newcommand{\tabref}[1]{Table~\ref{#1}}

\newcommand{\JOut}{J_\mathrm{out}}

\newcommand{\cs}[1][]{\ifthenelse{\equal{#1}{}}{{\muline{c}}}{\muline{c}^{(#1)}}}
\newcommand{\sigmaF}{\sigma^\rightarrow}
\newcommand{\sigmaB}{\sigma^\leftarrow}
\newcommand{\sF}{s_\rightarrow}
\newcommand{\sB}{s_\leftarrow}
\newcommand{\kF}{k_\rightarrow}
\newcommand{\kB}{k_\leftarrow}
\newcommand{\sP}{s_\mathrm{p}}
\newcommand{\sA}{s_\mathrm{a}}
\newcommand{\kP}{k_\mathrm{p}}
\newcommand{\kA}{k_\mathrm{a}}
\newcommand{\eps}{\varepsilon}
\newcommand{\cBase}{c^{(0)}}
\newcommand{\cBaseIn}{\cBase_\mathrm{in}}
\newcommand{\cBaseOut}{\cBase_\mathrm{out}}
\newcommand{\cEq}{c^\mathrm{eq}}
\newcommand{\cEqIn}{\cEq_\mathrm{in}}
\newcommand{\cEqOut}{\cEq_\mathrm{out}}
\newcommand{\phiBase}{\phi^{(0)}}
\newcommand{\phiEq}{\phi^\mathrm{eq}}

\newcommand{\phiOut}{\phi_\mathrm{out}}
\newcommand{\phiBaseIn}{\phiBase_\mathrm{in}}
\newcommand{\phiBaseOut}{\phiBase_\mathrm{out}}
\newcommand{\phiEqIn}{\phiEq_\mathrm{in}}
\newcommand{\phiEqOut}{\phiEq_\mathrm{out}}
\newcommand{\GammaIn}{\Gamma_\mathrm{in}}
\newcommand{\GammaOut}{\Gamma_\mathrm{out}}

\newcommand{\kOut}{k_\mathrm{out}}
\newcommand{\JIn}{J_\mathrm{in}}

\newcommand{\DOut}{D_\mathrm{out}}
\newcommand{\lOut}{l_\mathrm{out}}

\newcommand{\p}[1][]{\ifthenelse{\equal{#1}{}}{{\phi}}{\phi^{(#1)}}%
}
\newcommand{\m}[1][]{\ifthenelse{\equal{#1}{}}{{\mu}}{\mu^{(#1)}}%
}
\newcommand{\phiS}[1][]{\ifthenelse{\equal{#1}{}}{{\phi_\mathrm{S}}}{\phi_\mathrm{S}^{(#1)}}%
}

\title{Chemically active droplets}
\subtitle{Lecture Notes for the Boulder Summer School on Self-Organizing Matter}
\author{David Zwicker}

\date{Max Planck Institute for Dynamics and Self-Organization\\[2ex] \today}

\begin{document}
\maketitle
\begin{abstract}
These lecture notes describe a basic theory of chemically active droplets, which are droplets kept away from equilibrium by driven chemical reactions.
The notes assume a basic familiarity with equilibrium thermodynamics of  phase separation, and thus focus on three separate themes, which were discussed in three separate lectures:
(i) The kinetics of phase separation, including the early-stage dynamics of spinodal decomposition and the late-stage dynamics of Ostwald ripening.
(ii) Transition state theory as a simple, thermodynamically-consistent kinetic theory of chemical reactions, which permits explicit driving in open systems.
(iii) The combination of phase separation and reactions, leading to active droplets.
We discuss the two fundamental classes of internally-maintained and externally-maintained droplets.
A simple version of externally-maintained droplets permits an effective electrostatic analogy, which indicates how the reaction-diffusion system mediates long-ranged interactions.
All these aspects are discussed in the context of biomolecular condensates.
\end{abstract}

\clearpage
\tableofcontents
\clearpage

\section{Motivation}
Droplets formed by phase separation are ubiquitous in daily experiences, soft matter physics, and biology.
Phase separation is a basic physical process that can explain how material concentrates in a spatial region from passive interactions alone.
Chemical reactions are another omnipresent phenomenon, which is particularly important in engineering and biology.
Reactions that consume fuel molecules and produce waste are the basis of metabolism, and prevent life from reaching life-less thermodynamic equilibrium.
It is thus natural to wonder how chemical reactions affect droplets formed by phase separation, which is the topic of these lectures.

\subsection{Chemically active droplets in biology}
The fact that biological cells exploit phase separation to control the spatiotemporal organization of their biomolecules is currently revolutionizing cell biology.
The resulting intracellular droplets, which are also known as \emph{biomolecular condensates}, are relevant in all clades of life, from bacteria~\cite{Azaldegui2020}, over plants~\cite{Kim2021}, to animals~\cite{Banani2017}.
They are implicated in homeostasis of healthy cells (e.g., during development~\cite{So2021}, in gene regulation~\cite{Hirose2022}, and for signaling~\cite{Jaqaman2021}), but also in diseased states~\cite{Alberti2021} (e.g., neurodegenerative diseases~\cite{Mathieu2020}, and cancer~\cite{Cai2021}).
One central question is thus how cells regulate their condensates: \textbf{How do cells control where and when condensates form, how large they get, and when they disappear?}

It is likely that cells combine multiple physical processes to achieve such control.
For instance, simply producing or degrading the involved molecules surely gives control over droplet formation.
Similarly, global parameters, like salt concentrations or pH will affect phase separation and can be exploited for control.
Moreover, the immediate surrounding is also relevant, and heterogeneities in material properties or gradients of any of the parameters above will affect droplets.
An interesting additional alternative are direct chemical modifications of the involved biomolecules, e.g., phosphorylation of proteins or methylation of nucleic acids like RNA and DNA.
Such modifications clearly affect the interactions of those molecules with their surrounding and can thus be used to influence the process of phase separation.
In fact, there are many examples of biomolecular condensates that are regulated by such molecular modifications~\cite{Soeding2019,Snead2019,Hofweber2019}, including centrosomes~\cite{Zwicker2014} and stress granules~\cite{Hondele2019}.
These examples show that chemical reactions can affect phase separation, but the converse is also true:
Chemical reactions depend on the local chemical environment and will thus generally behave differently inside a condensate compared to the surrounding phase.
It is thus clear that condensation and reactions influence each other and we need to build a unified theory, based on thermodynamic arguments, to unravel this interplay.

\subsection{Phase separation and chemical reactions are linked}
To illustrate that phase separation and chemical reactions are intimately linked, let us first consider a simple equilibrium system comprising two species $A$ and $B$, which can both convert into each other, $A \rightleftharpoons B$, and also phase separate from each other~\cite{Zwicker2022a}.
Due to phase separation, there will be two different regions (or phases) in the system, where $A$ and $B$ assume different concentrations.
We can thus characterize the composition of the system by four concentrations: $c_A^{(1)}$, $c_A^{(2)}$, $c_B^{(1)}$, and $c_B^{(2)}$, where $(1)$ and $(2)$ denote the different phases.
The partitioning of the two species between the phases is characterized by \emph{partition coefficients},
\begin{align}
	\label{eqn:partition_coefficient_simple}
	P_A &= \frac{c_A^{(1)}}{c_A^{(2)}}
& \text{and} &&
	P_B &= \frac{c_B^{(1)}}{c_B^{(2)}}
	\;,
\end{align}
which can in principle be measured experimentally.
In contrast, the chemical equilibrium of the reaction $A \rightleftharpoons B$ is characterized by \emph{equilibrium constants},
\begin{align}
	\label{eqn:reaction_equilibrium_ratio_simple}
	K^{(1)} &= \frac{c_B^{(1)}}{c_A^{(1)}}
& \text{and} &&
	K^{(2)} &= \frac{c_B^{(2)}}{c_A^{(2)}}
	\;,
\end{align}
which can in principle differ between the two phases since the chemical environment is different.
Combining \Eqsref{eqn:partition_coefficient_simple} and \eqref{eqn:reaction_equilibrium_ratio_simple} implies
\begin{align}
	\label{eqn:relationship_simple}
	\frac{P_A}{P_B} = \frac{K^{(2)}}{K^{(1)}}
	\;.
\end{align}
Without phase separation, we necessarily have $P_A = P_B = 1$, implying $K^{(2)}=K^{(1)}$ for consistency.
However, this picture changes when phase separation takes place.
In particular if the two components partition differently ($P_A\neq P_B$), we necessarily have different equilibrium constants in the two phases!
This shows that the phase equilibrium (described by the partition coefficients) and the chemical equilibrium (described by the equilibrium constants) are directly linked.
This connection does not only affect thermodynamic equilibrium, but also the kinetics.
These kinetics will be particularly important when we consider active systems, which are kept away from thermodynamic equilibrium.
To study such systems, we next introduce the kinetics of phase separation, chemical reactions, and the combined, driven system, inspired by the review~\cite{Zwicker2022a}.

\clearpage
\section{Kinetics of phase separation}
\label{sec:phase_separation}

\subsection{Dynamics of conserved fields: Cahn-Hilliard equation}
We start by consider a system of a fixed volume~$V$ and temperature $T$ comprising a phase separating droplet material and a solvent.
The state of the system is characterized by the concentration field~$c(\vect r, t)$ of the droplet material, whereas the solvent fills the remaining space. 
Since material can only be redistributed, but not created nor destroyed, $c$ obeys a continuity equation
\begin{align}
	\partial_t c + \nabla.\vect j &= 0
	\;,
\end{align}
where $\vect j$ is a vector field describing the local flux of material.
In a passive system without hydrodynamic flows, this flux is driven by thermodynamic forces, i.e., the tendency of the system to lower the overall free energy.
In particular, differences in the chemical potential~$\mu$ in nearby positions will lead to a redistribution of material since moving a particle from a region of high chemical potential to one of lower chemical potential leads to an overall energy reduction.
In the simplest cases, when the system is not too far away from equilibrium, the induced flux is proportional to the change in energy,
\begin{align}
	\label{eqn:diffusive_flux}
	\vect j &= -\Lambda \nabla \mu
	\;,
\end{align}
where $\Lambda$ is a positive mobility coefficient, which captures kinetic details of the system.
Note that $\Lambda$ often depends on composition: in a dilute system, we generally assume $\Lambda \propto c$, because the flux ought to be proportional to the amount of material present.
Taken together, we find
\begin{align}
	\label{eqn:diffusion}
	\partial_t c &= \nabla.(\Lambda \nabla \mu)
	\;,
\end{align}
which is also known as model B~\cite{Hohenberg1977}.
The equation is a non-linear partial differential equation, which describes how material moves in the system to lower the overall free energy, which we need to specify to analyze the equation in detail.

\paragraph{Limit of ideal mixture}
In the simple case of an ideal fluid, we have $\mu = \kBT \ln c + w$ with constant reference chemical potential $w$, and we choose $\Lambda = \lambda c$.
This implies
\begin{align}
	\label{eqn:diffusion_ideal}
	\partial_t c &=  \kBT\lambda \nabla.(c \nabla  \ln c)
\notag\\	&=  \kBT\lambda \nabla.(c c^{-1} \nabla c)
\notag\\	&= D \nabla^2 c
\end{align}
with $D = \kBT \lambda$.
Consequently, we obtain Fick's law of ideal diffusion in this simple case.
However, \Eqref{eqn:diffusion} is valid much more generally, particularly when the chemical potential is a non-monotonous function of the composition, leading to phase separation.

\paragraph{Cahn-Hilliard equation}
A simple choice for the free energy~$F$ that can describe phase separation is given by the functional
\begin{align}
	\label{eqn:free_energy_functional_c}
	F[c] &= \int \left( f(c) + \frac{\kappa}{2}|\nabla c|^2\right) \diff V
	\;,
\end{align}
where the free energy density~$f(c)$ captures local interactions and is, e.g., given by Flory-Huggins theory~\cite{Flory1942, Huggins1941}, whereas the second term penalizes gradients, which will be important for describing interfaces~\cite{Cahn1958}.
Since the chemical potential is given by the functional derivative of  the free energy, $\mu = \delta F/\delta c$, we have
\begin{align}
	\label{eqn:chemical_potential_f}
	\mu = f'(c) - \kappa\nabla^2 c
	\;,
\end{align}
converting \Eqref{eqn:diffusion} to the general \emph{Cahn-Hilliard equation}
\begin{align}
	\label{eqn:cahn_hilliard}
	\partial_t c &= \nabla.\Bigl[\Lambda(c) \nabla \bigl(f'(c) - \kappa\nabla^2 c\bigr)\Bigr]
	\;,
\end{align}
which describes the dynamics of binary phase separating systems

\subsection{Early-stage dynamics: Spinodal decomposition}
\label{sec:spinodal_decomposition}

\begin{figure}[t]
	\centering	
	\includegraphics[width=0.7\textwidth]{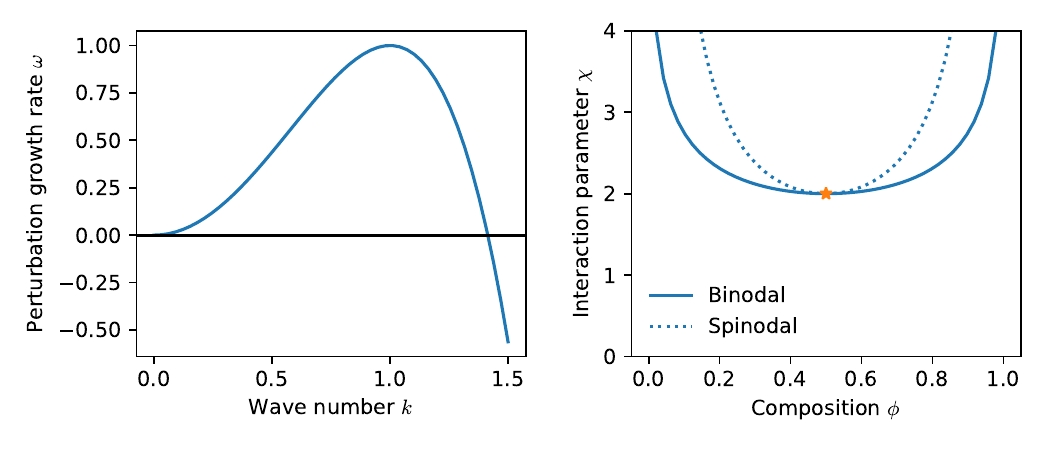}
	\caption{%
	\textbf{Instability of the homogeneous state.}
	Left: Perturbation growth rate $\omega$ as a function of wave number $k$ according to \Eqref{eqn:perturbation_growth_rate}.
	Right: Phase diagram illustrating the region where the homogeneous state has an instability (spinodal, dotted line) and the region in which a droplet is stable (binodal, solid line). The critical point is marked with a star.
	Analytical expressions are summarized in Appendix~\ref{app:binary_mixtures}.
	}
	\label{fig:spinodal}
\end{figure}

To understand how phase separation develops, we first focus on the early stage dynamics describing how droplets may form from a homogeneous state.
Each homogeneous state, $c(\vect r) = c_0$, is a stationary state of \Eqref{eqn:cahn_hilliard}, but such a state could be unstable, developing perturbations.
To see when this can be the case, we need to perform a stability analysis.
We thus consider $c(\vect r, t) = c_0 + \delta c(\vect r, t)$ for small perturbations $\delta c$, which we express using plane waves, $\delta c(\vect r, t)= \eps e^{\omega t + i \vect q.\vect r}$.
Up to linear order in $\eps$, we find from \Eqref{eqn:cahn_hilliard} that
\begin{align}
	\omega \delta c &= \nabla\Bigl(
		\bigl[\Lambda(c_0) + \Lambda'(c_0)\delta c\bigr]
		. \nabla \bigl[\cancel{f'(c_0)} + f''(c_0)\delta c - \kappa \nabla^2 \delta c\bigr] 
	\Bigr) \;,
\intertext{where the one term cancels because it does not depend on space. We can then also skip the term $\Lambda'(c_0)\delta c$ since it would lead to quadratic terms in $\delta c$, resulting in}
	\omega \delta c&= \Lambda(c_0)\nabla^2 \bigl[f''(c_0)\delta c - \kappa \nabla^2 \delta c\bigr] 
\notag\\
	&= -\Lambda(c_0)\vect q^2 \bigl[f''(c_0)\delta c + \kappa \vect q^2 \delta c\bigr] 
	\;.
\intertext{Dividing by $\delta c$, we thus obtain the perturbation growth rate}
	\label{eqn:perturbation_growth_rate}
	\omega &= -\Lambda(c_0)\vect q^2 \bigl[f''(c_0) + \kappa \vect q^2\bigr]
	\;,
\end{align}
which is the dispersion relation of this system; see \figref{fig:spinodal}.
The function $\omega(q)$ for $q=\abs{\vect q}$ tells us which wavelengths $\lambda = 2\pi/q$ are stable ($\omega < 0$) or unstable ($\omega > 0$).
Clearly, we have $\omega(0)=0$, which indicates that the overall amount of material does not change, consistent with the conservation law that led to  \Eqref{eqn:cahn_hilliard}.
We also find $\omega<0$ for small wavelengths (large $k$) since the interfacial term proportional to $\kappa$ suppresses such fluctuations.
Finally, \Eqref{eqn:perturbation_growth_rate} shows that intermediate wavelengths can only be unstable if $f''(c_0) < 0$, which is known as the \emph{spinodal region}; see \figref{fig:spinodal}.
In this case, small perturbations with a sufficiently large wavelength get amplified and grow exponentially with rate~$\omega$, which is known as \emph{spinodal decomposition}.
The wave vector with the largest growth rate $\omega$ is given by $q_\mathrm{max}^2 = -f''(c_0)/(2 \kappa)$.
We thus expect an unstable homogeneous system with concentration $c_0$ and some random fluctuations including all wavelengths (white noise) to first develop patterns with the spinodal wave length $\lambda_\mathrm{spin} = 2\pi/q_\mathrm{max}$,
\begin{align}
	\lambda_\mathrm{spin} &= 2\pi\sqrt{\frac{2\kappa}{-f''(c_0)}}
	\;.
\end{align}
However, this linear stability analysis can only predict the early-stage behavior, whereas later stages are governed by non-linear behavior of \Eqref{eqn:cahn_hilliard}.

\subsection{Late-state dynamics: Droplet coarsening by Ostwald ripening}

\begin{figure}[t]
	\centering	
	\includegraphics[width=0.5\textwidth]{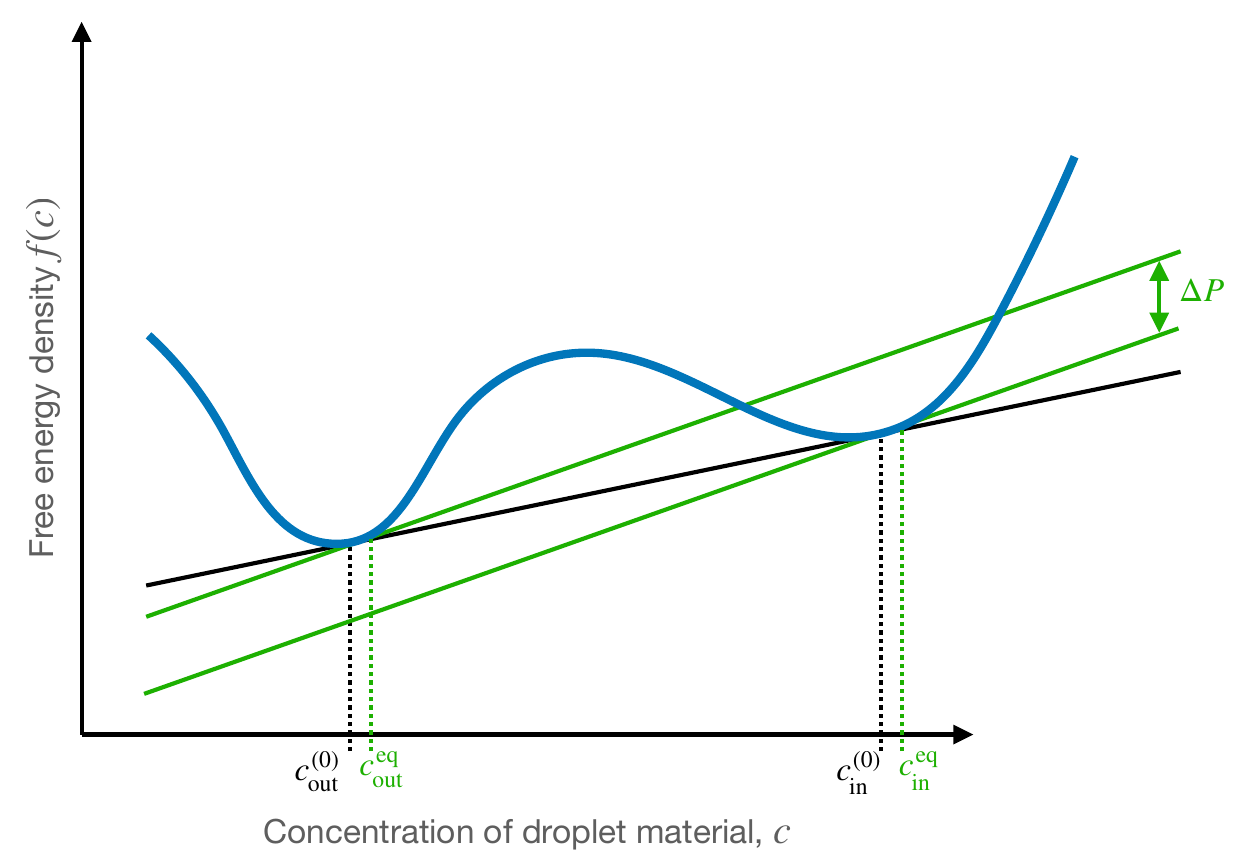}
	\caption{%
	\textbf{Maxwell construction obtains coexisting concentrations.}
	Free energy density $f$ as a function of the concentration~$c$ of the droplet material (blue line).
	The black lines indicate the Maxwell construction for thermodynamically large systems ($R\rightarrow\infty$, where Laplace pressure~$\Delta P$ is negligible). 
	The Maxwell construction is a common tangent with $f(c)$, so that the intersection points $\cBaseIn$ and $\cBaseOut$ satisfy the coexistence conditions given by \Eqsref{eqn:coexistence_conditions}.
	For finite droplets, the tangent splits into two (green lines), implying that coexisting concentrations $\cEqIn$ and $\cEqOut$ are elevated.
	}
	\label{fig:maxwell_construction}
\end{figure}

In the late stage, we have fully developed droplets, which are separated from the common dilute phase by a thin interface.
The width of this interface as well as the associated surface energy~$\gamma$ are controlled by the gradient term proportional to $\kappa$ in the free energy given by \Eqref{eqn:free_energy_functional_c}.
Both quantities generally scale with $\sqrt{\kappa}$, so a larger penalization of gradients leads to broader interfaces with a larger surface energy~\cite{Weber2019}, and the surface energy~$\gamma$ is equivalent to a surface tension~\cite{Kirkwood1949,Ip1994}.
The exchange of molecules across the interface is typically fast, so that the equilibrium conditions
\begin{subequations}
\label{eqn:coexistence_conditions}
\begin{align}
	\mu(\cEqIn) &= \mu(\cEqOut)
\\
	P(\cEqIn) &= P(\cEqOut) + \frac{2\gamma}{R}
	\label{eqn:pressure_balance}
\end{align}
\end{subequations}
are obeyed.
Similar to the chemical potential $\mu$, the osmotic pressure~$P$ of the solvent can also be determined from the free energy, $P=c\mu - f_\mathrm{full}$, where $\mu$ is given by \Eqref{eqn:chemical_potential_f} and $f_\mathrm{full} =f(c) + \frac12\kappa|\nabla c|^2$ is the integrand of \Eqref{eqn:free_energy_functional_c}~\cite{Zwicker2022a}.
The pressure balance given by \Eqref{eqn:pressure_balance} includes the Laplace pressure $\Delta P = 2\gamma/R$, which captures the effect of surface tension~$\gamma$ on a droplet of radius~$R$.
For a large droplet ($R\rightarrow\infty$), the interface is essentially flat and the pressure difference between the two phases vanishes.
In a binary mixture the associated coexisting concentration $\cBaseIn$ and $\cBaseOut$ are then controlled by the free energy density $f(c)$ alone, and can be determined from a \emph{Maxwell construction}; see \figref{fig:maxwell_construction}.
Based on these solutions, the correction caused by Laplace pressure~$\Delta P$ reads~\cite{Vidal2020}
\begin{align}
	\label{eqn:ceqout_from_pressure}
	\cEqOut(\Delta P) \approx \cBaseOut \exp\left(\frac{\Delta P}{\cBaseIn \kBT}\right)
	\;.
\end{align}
To linear order in $R^{-1}$, we find
\begin{align}
	\label{eqn:ceqout_from_radius}	
	\cEqOut(R) \approx \cBaseOut\left(1 + \frac{\ell}{R}\right)
	\;,
\end{align}
where $\ell = 2\gamma/(\cBaseIn \kBT)$ is sometimes called the \emph{capillary length}.
Consequently, Laplace pressure causes an elevation of the concentration outside droplets.
In particular, $\cEqOut$ is larger for smaller droplets, imposing a concentration gradient in the dilute phase between droplets of different size.
Since the dilute phase can be approximated as an ideal solution, these concentration differences cause a diffusive flux (see \Eqref{eqn:diffusion_ideal}), which transports material from small to larger droplets.
This material is supplied from the small droplet itself, so that it shrinks while the larger droplets grow.
Taken together, this diffusive transfer of material from small to large droplets leads to a coarsening of the entire emulsion, which is known as \emph{Ostwald ripening}.

\begin{figure}[t]
	\centering	
	\includegraphics[width=\textwidth]{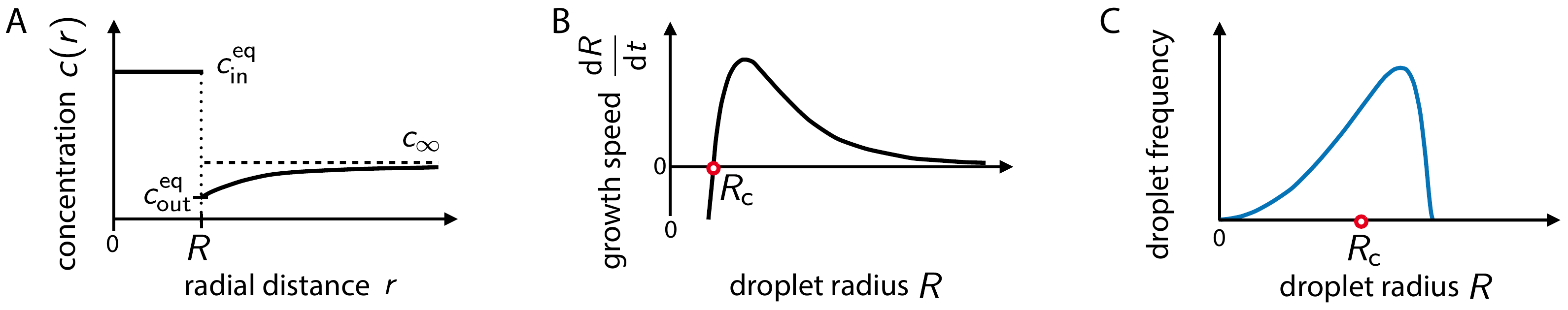}
	\caption{%
	\textbf{Dynamics of passive droplets.}
	\textbf{(A)} Concentration profile $c(r)$ around a single droplet of radius~$R$. Thermodynamic equilibrium at the interface sets the coexisting concentrations $\cEqIn$ and $\cEqOut$, which differs from the concentration~$c_\infty$ far away from the droplet.
	\textbf{(B)} Droplet growth rate as a function of $R$. Droplets grow if they are larger than the critical radius $\Rcrit$.
	\textbf{(C)} Universal size distribution in the late stage.
	(A--C) Figure adapted from \cite{Weber2019}.
	}
	\label{fig:ostwald_ripening}
\end{figure}

The detailed dynamics of Ostwald ripening can be understood by focusing on a single droplet of radius~$R$, which coexists with a dilute phase at an average concentration $c_\infty$; see \figref{fig:ostwald_ripening}a.
This average concentration of the dilute phase might slowly evolve with time, and generally differs from the concentration $\cEqOut(R)$ right outside the droplet.
The concentration difference drives a diffusive flux and determines whether droplets grow or shrink.
The droplet volume~$V$ shrinks when there is a net efflux~$\JOut$ of material away from the droplet,
\begin{align}
	\label{eqn:droplet_growth_passive}
	\partial_t V &= \frac{-\JOut}{\cEqIn - \cEqOut}
	\;.
\end{align}
The denominator in this equation denotes the concentration difference that needs to be "filled up" by the influx $-\JOut$ to increase the volume by an infinitesimal amount $\partial_t V$.
The efflux $\JOut$ can be estimated by solving the stationary diffusion problem, $\nabla^2 c=0$,  between the equilibrium concentration, $c(R) = \cEqOut$ and $c(r\rightarrow\infty)=c_\infty$, which leads to
\begin{align}
	c(r) &= c_\infty + (\cEqOut - c_\infty) \frac{R}{r}
	\;.
\end{align}
Combining $\JOut=4\pi R^2 j_\mathrm{out}$ with $j_\mathrm{out} = -D\partial_r c$ then implies
\begin{align}
	\label{eqn:flux_outside}
	\JOut \approx 4\pi R D \bigl(\cEqOut - c_\infty\bigr)
	\;.
\end{align}
Taken together, we thus find
\begin{align}
	\partial_t V \approx \frac{4\pi R D}{\cBaseIn - \cBaseOut}\left(c_\infty - \cBaseOut - \frac{\cBaseOut\ell}{R}\right)
	\;.
\end{align}
The dynamics of the droplet radius~$R$ can thus be summarized as
\begin{align}
	\label{eqn:dynamics_single_droplet}
	\partial_t R \approx \frac{D}{R(t)}\left( \Delta(t) - \frac{\alpha}{R(t)}\right)
	\;,
\end{align}
where $\alpha =\ell \cBaseOut/(\cBaseIn - \cBaseOut)$ and we have  the \emph{supersaturation} $\Delta = (c_\infty - \cBaseOut)/(\cBaseIn - \cBaseOut)$, which can be determined from material conservation,
\begin{align}
	N_\mathrm{tot} &= c_\infty \Vsys + \left(\cBaseIn - c_\infty\right) \sum_{i=1}^N V_i
	\;,
\end{align}
where $N_\mathrm{tot}$ is the total number of molecules of droplet material, $\Vsys$ is the system volume, and $V_i$ are the volumes of all $N$ droplets.
The dynamical \Eqref{eqn:dynamics_single_droplet} and \figref{fig:ostwald_ripening}b show that droplets grow if they are larger than the critical radius $\Rcrit = \alpha/\Delta$, which reads
\begin{align}
	\label{eqn:critical_radius}
	\Rcrit = \frac{	\ell \cBaseOut}{c_\infty - \cBaseOut}
	\;.
\end{align}
This critical size increases with decreasing super-saturation~$\Delta$, implying that more and more droplets disappear while the remaining ones grow in size.
This simple calculation thus explains Ostwald ripening qualitatively.

To obtain a rough idea of the quantitative dynamics of Ostwald ripening, we next focus on the late-stage regime, where the bracket in \Eqref{eqn:dynamics_single_droplet} is small.
In this case, we can assume that $\Delta \sim R^{-1}$, where $R$ now is the radius of a characteristic droplet in the system.
We thus find $\partial_t R \sim R^{-2}$, which implies $R(t) \sim t^{1/3}$.
This simple scaling analysis predicts that the typical droplet volume increases linearly with time, implying that the number of droplets decreases linearly.

\paragraph{Lifshitz-Slyozov-Wagner theory}
In a more quantitative treatment, one characterizes a polydisperse mixture, where many different droplet sizes coexist, by the droplet size distribution $P(R)$.
Lifshitz and Slyozov~\cite{Lifshitz1961} as well as Wagner \cite{Wagner1961} showed that this size distribution in the long-time limit attains a universal shape,
\begin{align}
	\hat P(\rho) &= \frac49 \rho^2 \left(1 + \frac{\rho}{3}\right)^{-\frac73}\left(1 - \frac{2\rho}{3}\right)^{-\frac{11}{3}}\exp\left(1 - \frac{3}{3-2\rho}\right)
	\;,
\end{align}
where $\rho=R/\Rcrit$ is the scaled radius; see \figref{fig:ostwald_ripening}c.
More importantly, the mean radius, which in this case is equal to the critical radius ($\mean{R} = \Rcrit$), scales as
\begin{align}
	\mean{R} &\propto \left(\frac{\ell D \cBaseOut}{\cBaseIn} t\right)^{\frac13}
	\;.
\end{align}
We again find that the average droplet radius increases as $t^{1/3}$. %
This effect is faster for larger surface tensions (implying larger $\ell$), larger diffusivities $D$,  and weaker phase separation (smaller $\cBaseOut/\cBaseIn$).

\subsection{Conclusions}
The dynamics derived here for binary systems also generalized to passive multicomponent systems~\cite{Zwicker2022a}.
In essence, starting from a homogeneous state, droplets first form either by stochastic nucleation by thermal noise (if the homogeneous state is linearly stable) or via spinodal decomposition.
The resulting droplets are endowed with a surface tension (which is an unavoidable consequence of phase separation), implying a spherical shape.
Moreover, this surface tension drives Ostwald ripening (and favors droplet coalescence), leading to fewer and larger droplets over time.
This coarsening process is a direct consequence of passive phase separation, and will thus be present in virtually all systems.
Although it might be slow (and thus insignificant) in some systems, a full control over droplet dynamics requires non-equilibrium processes.
We here focus on driven chemical reactions, but before we combine these with phase separation, we will first discuss chemical reactions in a well-mixed container.

\clearpage
\section{Chemical reactions}
\label{sec:chemical_reactions}

To derive basic kinetic equations for chemical reactions, we start by considering a well-mixed, homogeneous, isothermal system of fixed volume~$V$ that contains $N$ types of particles, which can in principle undergo chemical reactions.
The system's composition is fully described by the number concentrations $c_i = N_i/V$, where $N_i$ is the number of particles of type~$i$.
The behavior of the mixture is governed by a free energy density $f(\cs)$, which depends on the entire composition $\cs=\{c_1, c_2, \ldots, c_N\}$.
Crucial are the chemical potentials~$\mu_i = \partial f/\partial c_i$, which follow from the free energy density $f(\cs)$.
They can be expressed as
\begin{align}
	\label{eqn:chemical_potential}
	\mu_i &= \kBT \bigl[\ln c_i + w_i(\cs)\bigr]
	\;,
\end{align}
where the first term captures the translational entropy, whereas all other contributions are summarized by the (non-dimensional) enthalpic contributions $w_i(\cs)$, which can in principle depend on the entire composition $\cs$.
Note that $w_i$ is constant in an ideal solution, where it can be interpreted as the standard chemical potential, or an internal energy per molecule.
An expression alternative to \Eqref{eqn:chemical_potential} is $\mu_i = \mu_i^{0} + \kBT\ln(\gamma_i c_i)$, where $\mu^0_i$ is the constant reference chemical potential whereas the activity coefficients $\gamma_i(\cs)$ capture the non-ideality of the solution~\cite{Bauermann2021}.
Comparing the two equations, we find $w_i(\cs) = \ln\gamma_i(\cs) + \mu_i^{0}/\kBT$, and both formulations capture the same essence.

\subsection{Reaction equilibrium}
Let us start with a simple conversion reaction $A\rightleftharpoons B$ between two species $A$ and $B$.
Since $A$ can be converted into $B$ (and vice versa), the numbers of particles, $N_A$ and $N_B$, are no longer conserved individually.
However, the chemical conversion implies that the total particle count, $N_A + N_B$, stays constant.
To fully describe the equilibrium composition of the system, we thus additionally need to determine the ratio $N_B/N_A = c_B/c_A$, which is also known as the \emph{equilibrium constant} $K = c_B/c_A$.
It is determined by the chemical equilibrium, which is governed by the equivalence of chemical potentials, $\mu_A = \mu_B$.
The intuition is that if the chemical potentials are unbalanced, say $\mu_A$ is larger than $\mu_B$, the conversion can lower the overall free energy, in this case by converting $A \rightarrow B$.
For an ideal solution, where $w_A$ and $w_B$ are two constants, we find from \Eqref{eqn:chemical_potential} that $K = e^{w_A - w_B}$.
In this case, the ratio of the two components is fixed by the free energy and adding more material to the system only raises the total amount of both species.

Let us next consider a slightly more complicated reactions, where a molecule $A$ can combine with two molecules $B$ to form a complex~$C$, $2A + B \rightleftharpoons C$.
Note that the number of molecules created when the reaction runs once from left to right, summarized by the stoichiometry vector~$\sigma_i=(-2,-1,1)$, is less trivial than in the case above.
Consequently, the conserved quantity is less obvious, and possible choices include $N_A + N_B  + 3 N_C$, $\frac12 N_A + N_B + 2 N_C$, and $N_A + 2N_B + 4N_C$.
The pre-factors~$q_i$, here $(1,1,3)$, $(\frac12, 1, 2)$, and $(1, 2, 4)$, have in common that they obey $\sum_i q_i \sigma_i=0$, which ensures material conservation.
More importantly, each reaction only imposes one conservation law.
The remaining degrees of freedom, the particle numbers or concentrations of all but one species, in equilibrium must also obey the chemical equilibrium,
$2\mu_A + \mu_B = \mu_C$.
In this case of three different species, the conservation law and chemical equilibrium only comprise two conditions, so the chemical equilibrium is not unique.
Instead, the initial amounts of materials also affect how much material of each species there will be in equilibrium.
For instance, in the extreme situation that there is only $A$ and neither $B$ nor $C$, it is clear that the system must remain in that state.
The chemical equilibrium can again be characterized by an equilibrium ratio $K = c_C/(c_A^2c_B)$.
For ideal mixtures, we find $K = e^{2w_A + w_B - w_C}$, which again shows that the internal energies determine the reaction balance.

We next generalize the examples above to a general reaction involving $N$ different components $X_i$ for $i=1, \ldots N$.
The chemical reactions can be written as
\begin{align}
	\label{eqn:general_reaction}
	\sum_{i=1}^N \sigmaF_i X_i \rightleftharpoons \sum_{i=1}^N \sigmaB_i X_i
	\;,
\end{align}
where the left hand side (arbitrarily) denotes all reactants, whereas the species on the right hand side are called products.
The $\sigma^{\rightleftharpoons}_i$ denote the respective stoichiometric coefficients, which determine how many molecules of each component~$i$ participate in a single reaction.
Introducing the net stoichiometric coefficient $\sigma_i = \sigmaB_i - \sigmaF_i$, we can write the reaction more succinctly as $\sum_i \sigma_i X_i \rightleftharpoons \varnothing$.
The stochiometric coefficients are not arbitrary since the chemical reactions needs to conserve mass, $\sum \sigma_i M_i = 0$, where $M_i$ is the molecular mass of component~$i$.
In the case of multiple possible reactions, we have a stoichiometric coefficient vector $\sigma_i^{(\alpha)}$ for each reaction~$\alpha$, which together form the stoichiometric matrix.
Similar to the examples above, each reaction is associated with a conserved quantity $\psi_\beta$, which can be expressed as $\psi_\beta = \sum_i q_i^{(\beta)}N_i$, where $q_i^{(\beta)}$ are linearly independent vectors in the cokernel of the stoichiometric matrix, $\sum_i q_i^{(\beta)} \sigma_i^{(\alpha)} = 0$~\cite{Avanzini2021}.
Moreover, each reaction has its own chemical equilibrium condition,
\begin{align}
	\label{eqn:chemical_equilibrium}
	\sum_{i=1}^N \sigmaF_i \mu_i &= \sum_{i=1}^N \sigmaB_i \mu_i
& \text{or} &&
	\sum_{i=1}^N \sigma_i \mu_i &= 0
	\;.
\end{align}
This implies a general reaction constant $K = \prod_i c_i^{\sigma_i}$.
Inserting \Eqref{eqn:chemical_potential}, we find
\begin{align}
	K(\cs) &=  \exp\left[-\sum_{i=1}^N \sigma_i w_i(\cs)\right]
	\;,
\end{align}
which reduces to a constant for ideal mixtures where $w_i$ are constants.
We summarize the equilibrium thermodynamics of various chemical reactions in \tabref{tab:chemical_equilibrium}.

\begin{table}
\begin{center}
\caption{Examples of different reactions, showing reaction scheme, an example for the associated conserved quantity, the equilibrium condition, and the resulting reaction constant~$K$. Note that $K$ can in principle depend on the composition~$\cs$ in non-ideal mixtures, where $w_i$ is a function of $\cs$.}
\begin{tabularx}{\textwidth}{llll}
\toprule
Reaction & Conserved quant. & Equilibrium & Reaction constant\\
\midrule
$A\rightleftharpoons B$  & $N_A+N_B$ & $\mu_A = \mu_B$   & $K = e^{w_A - w_B}$ \\
$A  + B\rightleftharpoons 2C$ & $N_A + N_B + N_C$ & $\mu_A + \mu_B	 = 2 \mu_C$   & $K = e^{w_A + w_B - 2w_C}$ \\
$2A  + B\rightleftharpoons C$ & $N_A + N_B + 3 N_C$ & $2\mu_A + \mu_B =  \mu_C$   & $K = e^{2w_A + w_B - w_C}$ \\\midrule
$\displaystyle\sum_i \sigmaF_i X_i \rightleftharpoons \sum_i \sigmaB_i X_i$ &
	$\displaystyle\sum_i q_i N_i$ &
	$\displaystyle\sum_i \sigma_i \mu_i = 0$ &
	$\displaystyle K =  e^{-\sum_i \sigma_i w_i}$ \\
\bottomrule
\end{tabularx}
\label{tab:chemical_equilibrium}
\end{center}
\end{table}%

\subsection{Kinetics of general reaction: Transition State Theory}
So far, we have considered the equilibrium conditions of chemical reactions, but to investigate interesting chemical dynamics, we also need to specify their kinetics. 
The kinetics can be summarized by the reaction flux~$s$, which is the net number of reactions taking place per unit time.
The concentrations of the species involved in the reaction~\ref{eqn:general_reaction} thus change according to
\begin{align}
	\partial_t N_i &= \sigma_i s
	\;,
\end{align}
and there could in principle be additional terms if species $i$ participates in additional chemical reactions.
Focusing on a single reaction, we can separate the net flux~$s$ in the forward flux $\sF$ and the backward flux $\sB$,
\begin{align}
	s = \sF - \sB
	\;.
\end{align}
The flux necessarily vanishes in chemical equilibrium, $\sF = \sB$.
It turns out that detailed balance, which states that microscopic transitions are in equilibrium with their reverse direction, demands the stronger condition  known as \emph{detailed balance of the rates}~\cite{Weber2019},
\begin{align}
	\label{eqn:detailed_balance_rates}
	\frac{\sF}{\sB} 
		&=\exp\left(-\beta\sum_{i=1}^N \sigma_i \mu_i\right)
	\;,
\end{align}
where $\beta = 1/\kBT$ denotes the inverse thermal energy.
This condition ensures that the overall reaction always runs in the direction that reduces the overall free energy of the system, in accordance with the second law of thermodynamics.
In equilibrium, the exponent on the right hand side vanishes according to \Eqref{eqn:chemical_equilibrium}, implying  $\sF = \sB$.
\Eqref{eqn:detailed_balance_rates} restricts the ratio of the forward to backward rate, but it does not specify the individual reaction fluxes $\sF$ and $\sB$.

\begin{figure}[t]
	\centering	
	\includegraphics[width=0.5\textwidth]{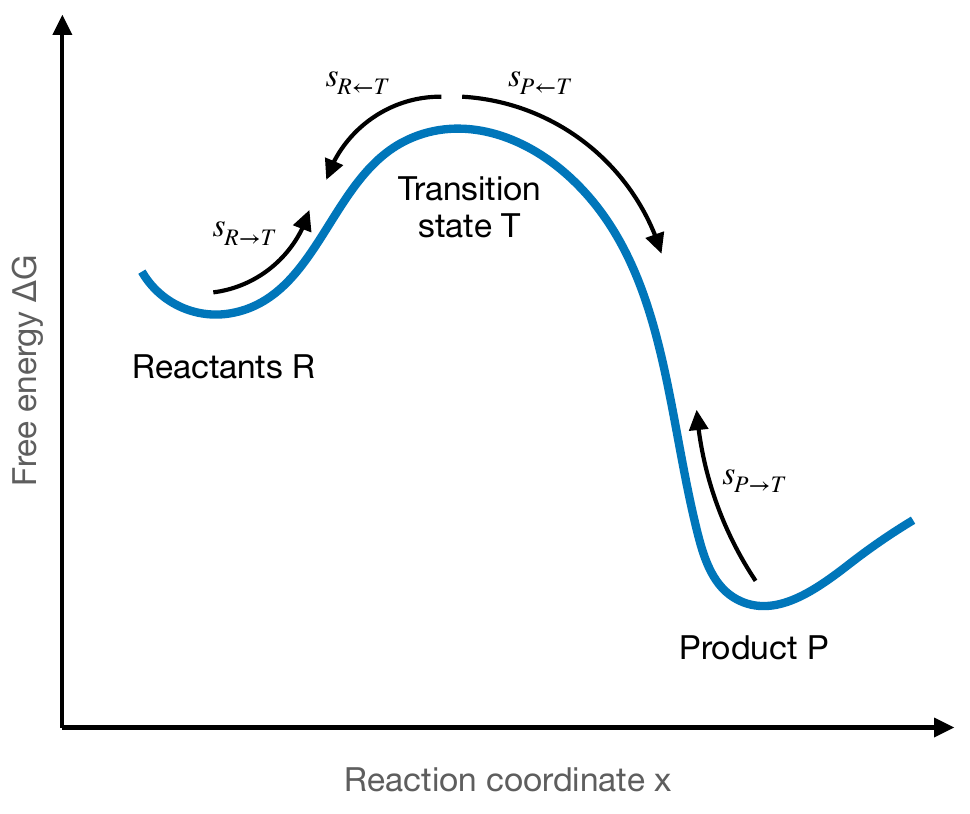}
	\caption{%
	\textbf{Transition state theory.}
	Schematic picture of the energy landscape (blue line) as a function of a suitable reaction coordinate for a reaction between reactants $R$ and products $P$ via a transition state $T$.
	}
	\label{fig:transition_state_theory}
\end{figure}

The exact details of the kinetics of a chemical reaction can be complicated, precisely because it is not solely governed by equilibrium thermodynamics.
To make progress, we instead need to make plausible assumptions to arrive at simple laws for chemical kinetics.
One of the simplest theories is \emph{transition state theory}, which assumes that the chemical reaction proceed by converting the reactants into an intermediate complex, known as the transition state, which is unstable and transforms into the product.
This progression of the chemical reaction can be best illustrated by an abstract reaction coordinate, which quantifies the progression from the products to the reactants; see \figref{fig:transition_state_theory}.
For a single reaction, we thus consider the initial reactants~$R$, the transitions state $T$, and the final product $P$, so that the entire reaction can be conceptualized as
\begin{align}
	\text{Reactants}~R \rightleftharpoons \text{Transition state}~T \rightleftharpoons \text{Products}~P
	\;.
\end{align}
The rates of the individual transitions are then derived under the assumption that an appropriate free energy can be associated with each of the three states.
In particular, detailed balance of the rates implies
\begin{align}
	\frac{s_{R \rightarrow T}}{s_{R \leftarrow T}} &=e^{\beta(\mu_R - \mu_T)}
& \text{and} &&
	\frac{s_{P \rightarrow T}}{s_{P \leftarrow T}} &=e^{\beta(\mu_P - \mu_T)}
	\;,
\end{align}
where $\mu_R$, $\mu_T$, and $\mu_P$ are the chemical potentials of the reactants, the transition state, and the products, so that their difference corresponds to the free energy change of the respective reaction.
The transition state~$T$ is unstable and thus quickly decays to either $R$ or $P$.
We assume that the rate~$s_T$ with which this happens is equal for both directions, $s_{R \leftarrow T} = s_{P \leftarrow T} = s_T$, which can be interpreted as a definition for the transition state itself.
We also assume that this decay is fast compared to the opposing directions, implying that the total fluxes are given by $\sF \approx \frac12 s_{R \rightarrow T}$ and $\sB \approx \frac12 s_{P \rightarrow T}$, which determine how often the transition state is reached from the reactants~$R$ and the products~$P$, respectively.
This condition implies that we assume a significant energy barrier between $R$ and $P$, $\mu_T - \mu_R > \kBT$ and $\mu_T - \mu_P > \kBT$.
Taken together, we thus arrive at
\begin{align}
	\sF &\approx \frac{s_T}{2} e^{\beta(\mu_R - \mu_T)}
&\text{and} &&
	\sB &\approx \frac{s_T}{2} e^{\beta(\mu_P - \mu_T)}
	\;.
\end{align}
Defining the constant $k = \frac12 s_T e^{-\beta\mu_T}$, this becomes
\begin{align}
	\sF &\approx k e^{\beta\mu_R}
&\text{and} &&
	\sB &\approx k e^{\beta\mu_P}
	\;,
\end{align}
showing that the kinetic details, given by the decay rate $s_T$, and the thermodynamics of the transition state, given by $\mu_T$, can all be combined in the single kinetic parameter~$k$.
Clearly, the overall rate decreases with larger energy barriers (higher $\mu_T$), but since $k$ also contains a complicated kinetic factor, we can simply take it as a kinetic coefficient, which needs to be determined from microscopic theories or experiments.

To use transition state theory for general reactions, we only need to specify the total chemical potential of the reactants and products,
\begin{align}
	\sF &= k e^{\beta\sum_i \sigmaF_i \mu_i}
&
	\sB &= k e^{\beta\sum_i \sigmaB_i \mu_i}
\end{align}
These naturally obey \Eqref{eqn:detailed_balance_rates}, and we obtain
\begin{align}
	\label{eqn:reaction_rate}
	s &= k\left(e^{\beta\sum_i \sigmaF_i \mu_i} - e^{\beta\sum_i \sigmaB_i \mu_i}\right)
\end{align}
for the total reaction flux of a general reaction.
Using the expressions for the chemical potentials given in \Eqref{eqn:chemical_potential}, we find
\begin{align}
	\label{eqn:reaction_rate_with_mu}
	s &= k\biggl( 
		e^{\sum_i \sigmaF_i w_i} \prod_j c_j^{\sigmaF_j}
	- e^{\sum_i \sigmaB_i  w_i} \prod_j c_j^{\sigmaB_j}
	\biggr)
	\;.
\end{align}
For ideal mixtures, $w_i$ are constants that do not depend on the composition, so we find
\begin{align}
	\label{eqn:mass_action_kinetics}
	s &= \kF \prod_j c_j^{\sigmaF_j} - \kB \prod_j c_j^{\sigmaB_j}
\end{align}
with $\kF = k e^{\sum_i \sigmaF_i w_i} $ and $\kB = k e^{\sum_i \sigmaB_i  w_i} $ defining the respective rate constants.
For the simple reaction $2A + B \rightleftharpoons C$ we discussed above, this results in $s = \kF c_A^2 c_B - \kB c_C$, which shows that \Eqref{eqn:mass_action_kinetics} describes simple mass action kinetics.
However, the more general equation \eqref{eqn:reaction_rate_with_mu} shows that these mass action kinetics can break down in interacting systems.

The general reaction rate described by \Eqref{eqn:reaction_rate} always proceeds in the direction that is thermodynamically favored, so that the system ends up in thermodynamic equilibrium after a long time, which is characterized by \Eqref{eqn:chemical_equilibrium}.
To describe actively driven systems, we next need to consider open systems, which allow matter exchange with the environment (beside the energy exchange already permitted by the coupled heat bath).

\subsection{Reactions driven away from equilibrium: Chemostating}
Open, non-equilibrium systems are a large class of physical systems that cannot all be described by the same simple theory.
We thus here restrict ourselves to simple cases, where the system contains fuel molecules~$F$ and waste molecules~$W$, which can be used to drive chemical reactions.
In contrast to the cases discussed so far, we assume that the concentration of the fuel and waste are kept constant, essentially because the system is allowed to exchange these particles with a large \emph{chemostat}.
Similar to the grandcanonical ensemble, this chemostat fixes the chemical potentials of the chemostatted species $F$ and $W$, so their total amount can in principle vary ($\mu$ ensemble).
In contrast, the other particles are conserved ($N$ ensemble), except for changes caused by the internal chemical reactions.
In biological systems, the fuel could correspond to ATP, whereas the waste would then be ADP and the single phosphate; An ATP replenishing system (like the mitochondria) then plays the role of the chemostat.
Similar chemical fuels could also apply to synthetic systems~\cite{Donau2023}, but the fuel could also correspond to photons provided by a light source, and the waste would then not be necessary.

To build intuition for driven reactions, we first consider the simple conversion 
\begin{align}
	A + F \rightleftharpoons B + W
	\;,
\end{align}
which now involves the fuel~$F$ and the waste~$W$.
The associated reaction rate is given by \Eqref{eqn:reaction_rate} and reads
\begin{align}
	s &= k\left(e^{\beta(\mu_A + \mu_F)} - e^{\beta(\mu_B + \mu_W)}\right)
	\;,
\end{align}
where we introduced the chemical potentials of the fuel and waste, which are kept fixed by the chemostat.
Hence, we can write this as
\begin{align}
	s &= \tilde k\left(e^{\beta(\mu_A + \Delta\mu)} - e^{\beta\mu_B}\right)
	\;,
\end{align}
where $\Delta\mu = \mu_F - \mu_W$ is the energy used in each reaction step and $\tilde k = k e^{\beta\mu_W}$ is a rescaled reaction rate.
Note that $\Delta\mu$ corresponds exactly to the energy extracted from the chemostat, which thus in principle allows us to quantify the rate of energy consumption (which must equal the entropy production rate in stationary state).

This theory can also be generalized to multiple reactions and an arbitrary number of chemostatted species, where extra care needs to be taken for reaction cycles~\cite{Avanzini2021}.
In our case, we consider the simpler situation where we only have one type of fuel and one type of waste, and there are no reaction cycles.
We can then generalize \Eqref{eqn:reaction_rate} to include the driving,
\begin{align}
	\label{eqn:driven_reaction}
	s &= k\left(e^{\beta(\tilde\sigma \Delta \mu + \sum_i \sigmaF_i \mu_i)} - e^{\beta\sum_i \sigmaB_i \mu_i}\right)
	\;,
\end{align}
where $\tilde\sigma$ is the number of fuel molecules  used (and thus also the number of waste molecules produced) during one reaction in the forward direction.

\subsection{Conclusions}
Chemical reactions break material conservation of the involved species, but each reaction still introduces a conserved quantity.
Note that some conservation laws might be broken by chemosatting~\cite{Avanzini2021}.
For example, for the two reactions $A \rightleftharpoons B$ and $A + F \rightleftharpoons B + W$, the sum $c_A + c_B$ is the only conserved quantity.
The rates of chemical reactions are defined by a combination of thermodynamics (encoded by the detailed balance of the rates, \Eqref{eqn:detailed_balance_rates}) and kinetics, similar to the diffusive dynamics in phase separation; see \Eqref{eqn:diffusive_flux}.
Using transition state theory, we can give explicit expressions, where the chemical potentials of the reactants and products enter exponentially, and the kinetics (as well as the thermodynamics of the transition state) are captured by a single kinetic pre-factor; see \Eqref{eqn:reaction_rate}.
Driving the system away from equilibrium is then straight-forward by simply keeping the chemical potentials of a fuel and a waste component constant.

\clearpage
\section{Phase separation with chemical reactions}
We finally combine the physics of phase separation discussed in \secref{sec:phase_separation} with (driven) chemical reactions discussed in \secref{sec:chemical_reactions}.

\subsection{Combination of phase separation and chemical reactions}
Phase separation plays out in space, so we now need to discuss concentration fields and can no longer assume that our system is well-mixed.
However, chemical reactions are local, so we can use the results, and particularly \Eqref{eqn:driven_reaction}, for each point in space.
Consequently, we describe a system comprising $N$ solutes and a solvent by $N$ volume fraction fields $\phi_i(\vect r, t)$, so the solvent fraction is $\phi_0 = 1-\sum_i\phi_i$.
For simplicity, we assume an incompressible system, so the molecular volumes~$\nu_i$ of all components are constant.
Consequently, the number concentrations $c_i$ that we used to define chemical reactions are proportional to the volume fractions, $c_i = \phi_i/\nu_i=N_i/V$, and we have $\partial_t \phi_i = \nu_i \sigma_i s$,
where $s$ is now the reaction flux, quantifying the number of reactions per unit time and unit volume.
Transition state theory the implies
\begin{align}
	\label{eqn:driven_reaction_phi}
	s &= k\left(e^{\beta(\tilde\sigma \Delta \mu + \sum_i \sigmaF_i \bar\mu_i)} - e^{\beta\sum_i \sigmaB_i \bar\mu_i}\right)
	\;,
\end{align}
where we needed to introduced the exchange chemical potentials $\bar\mu_i$ to respect incompressibility.
In essence, $\bar\mu_i$ denotes the free energy increase of the system when a molecule of species~$i$ is inserted into the system while $\nu_i/\nu_0$ solvent molecules are removed.
Moreover, we still keep the energy input $\Delta\mu$ from converting $\tilde\sigma$ fuel molecules~$F$ into waste molecules~$W$, which will allow us to drive reactions out of equilibrium.
For simplicity, we assume that the chemostatted species $F$ and $W$ are dilute, do not interact with other species, and diffuse fast, implying that their chemical potentials are constant and homogeneous everywhere.

The combination of phase separation with chemical reactions provides a lot of freedom, which we are only beginning to understand.
As a motivating example, let us consider a three-component system of two solutes $A$ and $B$, together with solvent $S$.
We consider a situation where $B$ phase separates from $A$ and $S$, while $A$ and $B$ can be converted into each other using a passive and a driven reaction,
\begin{align}
	A &\rightleftharpoons B
&
	A  + W &\rightleftharpoons B + F
	\;.
\end{align}
We set up the system such that the soluble species~$A$ has higher chemical potential compared to $B$, so that the phase separating form $B$ is generated spontaneously.
In contrast, we use the fuel~$F$ to turn $B$ back into $A$, and we restrict this reaction to the droplet by increasing the kinetic pre-factor inside the droplet.
Taken together, the dynamics of this system is described by~\cite{Kirschbaum2021}
\begin{subequations}
\label{eqn:example_full_system}
\begin{align}
\partial_t \phi_A &= \Lambda_A \nabla^2 \bar\mu_A - s_\mathrm{p} - s_\mathrm{a}
&
s_\mathrm{p} &= k_\mathrm{p}\left(e^{\beta\bar\mu_A} - e^{\beta\bar\mu_B} \right)
\\
\partial_t \phi_B &= \Lambda_B \nabla^2 \bar\mu_B + s_\mathrm{p} + s_\mathrm{a}
&
s_\mathrm{a} &= k_\mathrm{a}\phi_E\left(e^{\beta\bar\mu_A} - e^{\beta(\bar\mu_B + \Delta\bar\mu)} \right)
\;,
\end{align}
\end{subequations}
where the exchange chemical potentials $\bar\mu_i = \nu \delta F/\delta\phi_i$ are derived from the Flory-Huggins free energy
\begin{align}
	\label{eqn:example_full_system_free_energy}
	F[\phi_A, \phi_B] &= \frac{\kBT}{\nu} \int\left[
		\sum_{i=A,B,S} \!\!\!\! \phi_i\ln(\phi_i) 
		+ w\phi_A 
		+\!\!\!\! \sum_{i,j=A,B}\Bigl(\frac{\chi_{ij}}{2} \phi_i\phi_j + \frac{\kappa_{ij}}{2} (\nabla \phi_i).(\nabla\phi_j)\Bigr)
	\right]\diff V
	\;,
\end{align}
where we assumed that all species have equal molecular volume~$\nu$ and the solvent fraction reads $\phi_S = 1 - \phi_A - \phi_B$.
The linear term proportional to $w$ captures the higher internal energy of species $A$, whereas the interactions between the components are described by the Flory matrix~$\chi_{ij}$ and the interface coefficients~$\kappa_{ij}$.

Numerical simulations of \Eqsref{eqn:example_full_system}--\eqref{eqn:example_full_system_free_energy} show that droplets still form when the chemical reactions are comparably weak to phase separation.
However, the usual Ostwald ripening stops after a while, so that many droplets coexist and attain very similar radii.
This behavior is clearly caused by the driven chemical reactions that modify the phase separation behavior, and we call these systems \emph{chemically active droplets}.
To understand the behavior in more detail, we will next investigate how reactions affect phase separation, and we will then analyze the simplest possible model of active droplets.

\subsection{Passive binary mixture}
We now consider a binary system comprising two species $A$ and $B$, which can be converted into each other.
The state of the system is thus specified by the volume fraction~$\phi=\phi_B$ of species $B$, whereas $\phi_A = 1-\phi$.
Because of incompressibility, we only have one exchange chemical potential~$\bar\mu \propto \delta F/\delta \phi$.
Gradients of $\bar\mu$ drive relative diffusive fluxes between $A$ and $B$, whereas $\bar\mu$ itself drives reactions between the two species,
\begin{align}
	\label{eqn:dynamics_binary_passive}
	\partial_t \phi = \nabla \cdot \bigl[ \Lambda(\phi) \nabla \bar\mu\bigr] - \kP\bar\mu
	\;,
\end{align}
where we for simplicity linearized the reaction flux.
Note the negative sign in front of $\kP$ which ensures thermodynamic consistency.
\Eqref{eqn:dynamics_binary_passive}  describes a passive system, so its stationary state minimizes the associated free energy
\begin{align}
	\label{eqn:free_energy_functional}
	F[\phi] &= \frac{\kBT}{\nu} \int\left[
		f_0(\phi)
		+ \frac{\kappa}{2}|\nabla\phi|^2
	\right]\diff V
	\;.
\end{align}
Because the fractions of the individual species are no longer conserved, the system can attain the minimum of $f_0(\phi)$ at every point in space, implying a homogeneous system, so the gradient-term proportional to $\kappa$ also does not contribute.
In this case, there is clearly no diffusive flux, and the reaction flux also vanishes since $\bar\mu \propto f_0'(\phi)$ vanishes at the minimum.
Taken together, this shows that a binary system that allows for conversions between the two species always attains a homogeneous state in equilibrium.

In a system with more components, each chemical reaction reduced the number of independent components by one, increasing the degrees of freedom that the system can explorer.
Gibbs phase rule then implies that fewer phases can form.
For the binary system, we are left with only one phase, whereas a three-component system with one reaction can still separate into two phases.
However, such a system would still behave as a passive phase separating system, and particularly exhibit Ostwald ripening.
To break this behavior, we need to drive the system away from equilibrium.

\subsection{Active binary mixture}
We next consider a binary system that is driven out of equilibrium by the additional reaction $A + W \rightleftharpoons B + F$.
Again using linearized chemical reactions, we have
\begin{align}
	\partial_t \phi = \nabla \cdot \bigl[ \Lambda(\phi) \nabla \bar\mu\bigr] -\kP\bar\mu- \kA(\bar\mu - \Delta\mu)
	\;.
\end{align}
To get an intuitive understanding of the behavior of this system, we first investigate homogeneous states.
In particular, homogeneous states are stationary if the total reaction flux vanishes, which is the case if $\bar\mu =  \Delta\mu\kA/(\kA + \kP)$.
This clearly reduces to the passive case if $\Delta\mu = 0$, so we recover the results from the previous section.
In all other cases, the system is driven away from equilibrium, and both the passive and the active reaction have a sustained flux in stationary state, $\sA = -\sP =  \Delta\mu\kA\kP/(\kA + \kP)$.
The associated entropy production is simply $\sA\Delta\mu = \Delta\mu^2\kA\kP/(\kA + \kP)$.
Even though this active system is driven away from equilibrium and produces entropy continuously, stationary states are homogeneous if the coefficients $\kA$ and $\kP$ constants.
In essence, the chemical fluxes of the passive and active reaction balance at each position in space, converting material in \emph{futile cycles}, but not achieving any interesting spatial dynamics.

To convert the chemical energy injected via the active reaction into spatial fluxes, we need to break the symmetry of the reactions.
A simple solution is to demand that $\kA$ or $\kP$ depend on the composition~$\phi$, e.g., because the reactions are catalyzed by enzymes, fuel is segregated unequally, or the different chemical milieu affects the kinetics.
In this case, the fluxes $\sP$ and $\sA$ depend on $\phi$ and $\bar\mu$, allowing tremendous freedom in how to choose the specific kinetics.
One example for such reactions was already discussed above, in \Eqref{eqn:example_full_system}.
We will not study all possibilities here, but instead summarize the composition-dependent reactions by a simple rate law $s(\phi)$, so that we arrive at the dynamical equation
\begin{align}
	\label{eqn:dynamics_binary_rate_law}
	\partial_t \phi = \nabla \cdot \bigl[ \Lambda(\phi) \nabla \bar\mu\bigr] + s(\phi)
	\;.
\end{align}
Note that $s(\phi)$ now contains both the thermodynamic drive via the chemical potential $\bar\mu(\phi)$ as well as the composition-dependent kinetics.
The functional form is no longer restricted by thermodynamics since we explicitly consider an active system, but we learned from this section that interesting dynamics require a driven system with non-trivial kinetics.

\subsubsection{Stability of the homogeneous state}

\begin{figure}[t]
	\centering	
	\includegraphics[width=0.5\textwidth]{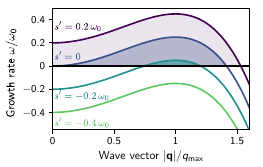}
	\caption{%
	\textbf{Stability analysis of homogeneous state in active binary system.}
	Perturbation growth rate $\omega$ as a function of wave vector $\vect{q}$ normalized using $q_\mathrm{max}^2 = -f''(\phi_*)/(2 \kappa)$ and $\omega_0 = \Lambda(\phi_*) [f''(\phi_*)]^2/(4\kappa)$ in the spinodal region where $f''(\phi_*)<0$.
	Figure adapted from \cite{Zwicker2022a}.
	}
	\label{fig:binary_active_stability}
\end{figure}

We start by analyzing the stability of the homogeneous state $\phi(\vect r) = \phi_*$, where the composition~$\phi_*$ of the homogeneous state satisfies $s(\phi_*)=0$.
Repeating the stability analysis from section \secref{sec:spinodal_decomposition}, we find
\begin{align}
	\label{}
	\omega &= -\Lambda(\phi_*)\vect q^2 \bigl[f_0''(\phi_*) + \kappa \vect q^2\bigr] + s'(\phi_*)
	\;.
\end{align}
Consequently, the only difference compared to passive phase separation is that the sensitivity $s'(\phi_*)$ provides an offset for the dispersion relation $\omega(q)$; see \Eqref{eqn:perturbation_growth_rate} and \figref{fig:binary_active_stability}.
In particular, the global mode $q=0$ is now unstable when $s'(\phi_*) > 0$.
This is not surprising since such a reaction is autocatalytic, where larger values of $\phi$ increase the production of $\phi$.
We will come back to such autocatalytic reactions below, but we first consider the opposite case, where $s'(\phi_*)<0$, which stabilizes the global mode ($q=0$).
If the reaction is strong, and $s'(\phi_*)$ is large negative, all $q$-modes can be stabilized, implying that the system stays in a homogeneous state.
Consequently, active chemical reactions can suppress phase separation, even if the parameters of the passive phase separating system are in the spinodal regime.
In an intermediate regime, for weakly negative $s'(\phi_*)$, only a band of modes is unstable, similar to the seminal Turing patterns~\cite{Cross2009}.
This behavior already suggests that the system selects a particular length scale, but to analyze this in detail, we need to investigate the non-linear behavior of the system.

\subsubsection{Effective droplet model: Linearization in both phases}
\begin{figure}[t]
	\centering	
	\includegraphics[width=\textwidth]{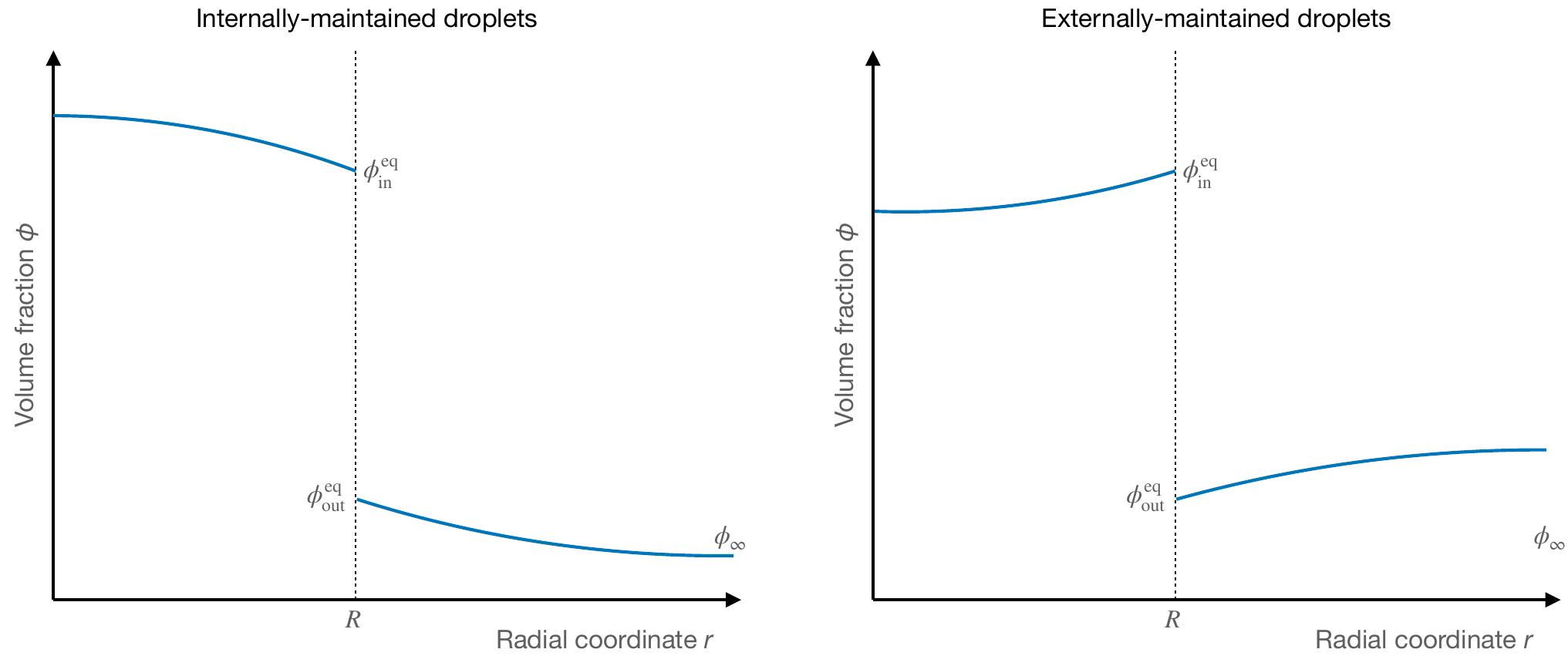}
	\caption{%
	\textbf{Concentration profiles of stationary active droplets.}
	Volume fraction $\phi$ of droplet material~$B$ as a function of the radial distance~$r$ from the center of the droplet.
	The fractions are controlled by the phase equilibrium at the interface at $r=R$ and maintain gradients inside and outside the droplet because of the chemical reactions.
	These gradients are opposite in the two cases of active droplets that are compared in the two panels.
	}
	\label{fig:active_droplet_profiles}
\end{figure}

\begin{figure}[t]
	\centering	
	\includegraphics[width=\textwidth]{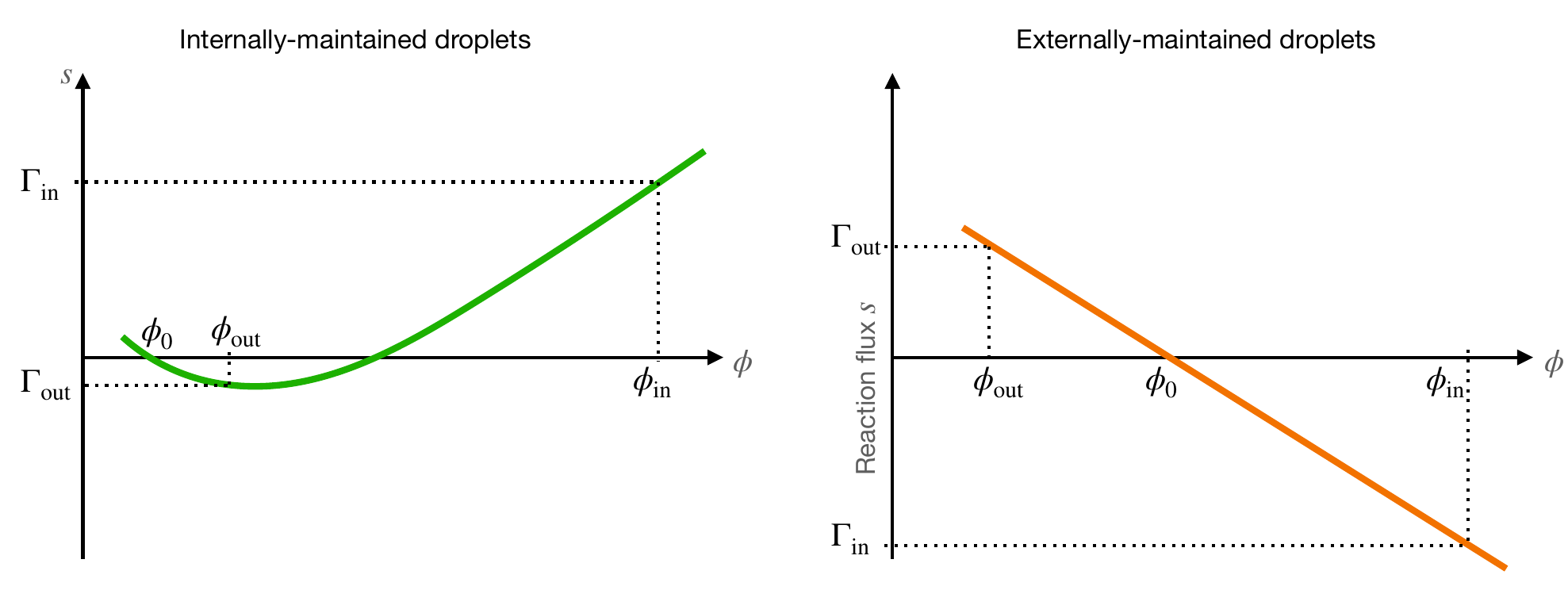}
	\caption{%
	\textbf{Examples for reaction rate laws for active droplets.}
	Reaction flux $s$ as a function of composition $\phi$ for the two classes of chemically active droplets:
	Internally-maintained droplets (left) exhibit an undersaturated dilute phase ($\phi_0<\phiOut$), degradation in the dilute phase ($\GammaOut<0$), and production in the dense phase ($\GammaIn>0$).
	In contrast, externally-maintained droplets exhibit a super-saturated dilute phase  ($\phi_0>\phiOut$), production in the dilute phase ($\GammaOut<0$), and degradation in the dense phase ($\GammaIn>0$).
	Note that a stable chemical equilibrium requires $s'(\phi)<0$, so the second root on the left hand side cannot be a homogeneous state.
	}
	\label{fig:active_droplet_reaction_rates}
\end{figure}

We next focus on the non-linear behavior in the late stage, when droplets have formed.
We thus consider weak chemical reactions, which do not prevent phase separation.
In this situation, we have well-formed droplets enriching the $B$ material, separated by a thin interface from the surrounding phase enriched in $A$.
The weak reactions are negligible at the interface, so the interfacial physics are dominated by phase separation.
In particular, we then know that right inside the interface (in the droplet phase), the $B$ fraction is given by $\phi=\phiEqIn$, whereas we have $\phi=\phiEqOut$ outside; see \Eqref{eqn:ceqout_from_radius}.
Away from the interface, in the bulk phases, chemical reactions can affect the volume fraction profile, but since reactions are weak, we expect these perturbations to be minor.
Consequently, we linearize the dynamics given by \Eqref{eqn:dynamics_binary_rate_law} in both phases around $\phi = \phiBase_\alpha$, where $\alpha \in \{\text{in}, \text{out}\}$.
Note that we linearized around the constants $\phiBase_\alpha$, since $\phiEq_\alpha$ contains minor surface tension effects, too.
The linearization results in 
\begin{align}
	\label{eqn:reaction_diffusion}
	\partial_t \phi &= D_\alpha \nabla^2 \phi + s_\alpha(\phi)
	\;,
\end{align}
where $D_\alpha = \Lambda(\phiBase_\alpha)f_0''(\phiBase_\alpha)$ denotes the diffusivity in the respective phase~$\alpha$.
The associated reaction flux reads
\begin{align}
	s_\alpha(\phi) &= \Gamma_\alpha - k_\alpha\bigl(\phi - \phiBase_\alpha\bigr)
	\;,
\end{align}
with the basal reaction flux $\Gamma_\alpha = s(\phiBase_\alpha)$ and the reaction sensitivity $k_\alpha = -s'(\phiBase_\alpha)$.
The basal reaction fluxes can be positive (so they produce droplet material~$B$) or negative (degradation of $B$).
If the basal fluxes were the same in both phases ($\GammaIn\GammaOut > 0$), droplet material $B$ would either be produced or degraded everywhere, which is inconsistent with stationary states.
Consequently, interesting dynamics can only take place when droplet material is produced in one phase while it is degraded in the other, so a circular flux of material can be sustained in the stationary state.
This naturally gives rise to two different classes of active droplets:
For \emph{internally-maintained droplets}, droplet material $B$ is produced inside the droplets ($\GammaIn > 0$) and must thus be degraded outside ($\GammaOut < 0$).
In contrast, \emph{externally-maintained droplets} produce $B$ outside ($\GammaOut>0$) and degrade it inside ($\GammaIn < 0$).
\figref{fig:active_droplet_profiles} shows the concentration profiles associated with these two cases.
Examples for reaction fluxes $s(\phi)$ for the two cases are shown in \figref{fig:active_droplet_reaction_rates}.

\begin{figure}[t]
	\centering	
	\includegraphics[width=0.6\textwidth]{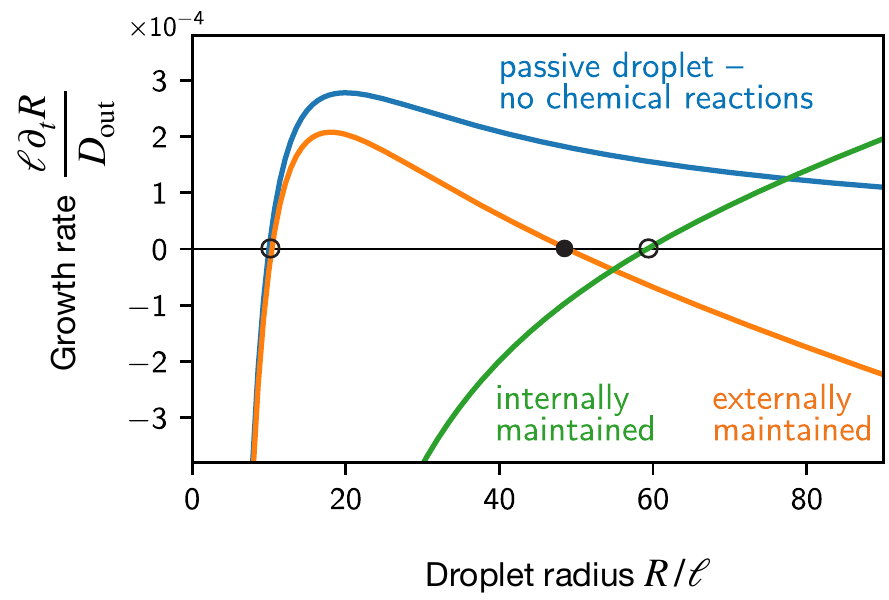}
	\caption{%
	\textbf{Growth rate of an active droplet.}
	Growth rate $\partial_t R$ as a function of droplet radius~$R$ for passive droplets (blue), internally-maintained droplets (green), and externally-maintained droplets (orange).
	Black disks and circles mark stable and unstable stationary states, respectively.
	Figure adapted from reference \cite{Weber2019}.
	}
	\label{fig:active_droplet_growth_rate}
\end{figure}

To understand internally-maintained and externally-maintained droplets more quantitatively, we next consider the simple situation of an isolated droplet in a large dilute phase.
Far away from the droplet the dilute phase is homogeneous, $\phi(\vect r)=\phi_\infty$, so the chemical reactions must balance, $s(\phi_\infty)=0$, implying
\begin{align}
	\phi_\infty = \phiBaseOut + \frac{\GammaOut}{\kOut}
	\;.
\end{align}
Solving for the stationary state of \Eqref{eqn:reaction_diffusion} outside the droplet, imposing boundary conditions $\phi(R) = \phiEqOut$ and $\phi(r\rightarrow \infty)=\phi_\infty$, leads to
\begin{align}
	\phi_\mathrm{out}(r) = \phi_\infty + (\phiEqOut-\phi_\infty) \frac{R}{r} \exp\left(\frac{R-r}{\lOut}\right) 
	\;,
\end{align}
with the reaction-diffusion length $\lOut=\sqrt{\DOut/\kOut}$.
Consequently, the integrated flux, $\JOut=4\pi R^2 j_\mathrm{out}$ with $j_\mathrm{out} = -\DOut\partial_r \phi_\mathrm{out}$, reads
\begin{align}
	\JOut \approx 4\pi \left(R + \frac{\lOut}{R}\right) \DOut \bigl(\phiEqOut - \phi_\infty\bigr)
	\;.
\end{align}
In the typical case of small droplets ($R\ll \lOut$) this expression is equivalent to \Eqref{eqn:flux_outside} for passive droplets.
The flux away from the droplet is positive if the dilute phase is undersaturated ($\phi_\infty < \phiEqOut$), which is the case when $\GammaOut < 0$, since $\kOut > 0$ for the homogeneous state to be stable ($s'(\phi_\infty) <0$; see previous section).
Consequently, internally-maintained droplets loose material since the degradation of droplet material in the dilute phase implies an undersaturation.
In contrast, externally-maintained droplets ($\GammaOut>0$) exhibit a super-saturated dilute phase ($\phi_\infty > \phiEqOut$), and gain material from the outside ($\JOut < 0$).
In a potential stationary state, the efflux $\JOut$ must be balanced by the flux~$\JIn$ created by the chemical reactions inside the droplet.
In the simplest case of a small droplet, the concentration profile inside the droplet is roughly given by $\phi(\vect r) \approx \phiBaseIn$, implying $\JIn \approx V \GammaIn$, where $V=\frac43\pi R^3$ is the droplet volume.
Given the two fluxes, we can directly determine the droplet growth rate,
\begin{align}
	\label{eqn:droplet_growth_active}
	\partial_t V &= \frac{\JIn -\JOut}{\phiEqIn - \phiEqOut}
	\;.
\end{align}
which is a straight-forward extension of \Eqref{eqn:droplet_growth_passive} for passive droplets.
\figref{fig:active_droplet_growth_rate} compares  droplet growth rates of the two active scenarios to the passive case.

To obtain more insight into the behavior, we next analyze \Eqref{eqn:droplet_growth_active} as a dynamical system~\cite{Strogatz2018}.
In particular, we can determine the stationary droplet radii radius $R_*$, which leads to
\begin{align}
	0 = \underbrace{\frac{\GammaIn}{3}}_{a}R_*^3 + \underbrace{\frac{\DOut \GammaOut}{\kOut} }_{b} R_* - \underbrace{ \DOut \lOut \phiBaseOut}_c
	\;.
\end{align}
The roots of the corresponding polynomial $0=aR_*^3 + bR_* - c$ can be determined in the limit of small droplets ($a\approx 0$) and large droplets ($c\approx 0$), leading to the two solutions
\begin{subequations}
\begin{align}
	R_{*,1} &\approx \frac{\ell}{\epsilon}
\\
	R_{*,2} & \approx \left(\frac{3\DOut\epsilon\phiBaseOut}{-\GammaIn}\right)^{\frac12} - \frac{\ell}{2\epsilon}
	\label{eqn:active_droplet_radius2}
	\;,
\end{align}
\end{subequations}
where we defined the supersaturation $\epsilon = \phi_\infty/\phiBaseOut - 1 = \GammaOut/(\kOut\phiBaseOut)$.
The third solution of the polynomial is always negative and thus unphysical.
Moreover, the first solution only provides positive values for $\epsilon > 0$, i.e., for externally-maintained droplets.
In contrast, the second solution, which also contains a linear correction for the surface tension term proportional to the capillary length~$\ell$,
is always real since $\epsilon\GammaIn < 0$ for both cases; The value of $R_{*,2}$ is typically positive unless reactions or surface tension are very strong.
A stability analysis reveals that $R_{*,1}$ is an unstable stationary state when it exists, whereas $R_{*,2}$ is unstable for internally-maintained droplets, whereas it is stable for externally-maintained droplets.
The behavior of these two classes of droplets is summarized in \tabref{tab:active_droplets}.

Internally-maintained droplets exhibit a large critical radius~$R_{*,2}$ since they exist in an under-saturated environment, and the efflux of material must be compensated by the production inside.
However, once they exceed this critical radius, they grow indefinitely, essentially because larger droplets exhibit a higher influx of droplet material compared to smaller droplets.
This results in an autocatalytic acceleration of growth until the internally-maintained droplet takes over the entire system.
The detailed dynamics need to be described by more detailed equations since the assumptions of a small droplet residing in a large dilute phase will eventually be violated.
Moreover, additional modifications, like a catalytic core and a limited amount of material can also alter the observed dynamics~\cite{Weber2019}.
Since internally-maintained droplets exist in an undersaturated environment, they don't nucleate spontaneously and instead can be formed at controlled locations by additional reactions~\cite{Soeding2019}.
One biological example for internally-maintained droplets are centrosomes~\cite{Zwicker2014}.

The behavior of externally-maintained droplets is qualitatively different compared to internally-maintained ones:
They exhibit two stationary states, of which the smaller one ($R_{*,1}$) is unstable and plays the role of the critical radius.
In fact, its value depends on the supersaturation~$\epsilon$ in exactly the same way as in passive droplets; see \Eqref{eqn:critical_radius}.
Similar to passive droplets, externally-maintained droplets thus grow spontaneously by taking up material from their surrounding if they exceed their critical radius.
However, in contrast to passive droplets, this growth stops when the active droplets reach the second, larger stationary radius ($R_{*,2}$), which is stable.
This theory predicts that  droplets that are larger than $R_{*,2}$ loose material (since the efflux due to degradation of droplet material in the droplet dominates the influx from the dilute phase) and thus shrink back.
Note that the first, dominant term in $R_{*,2}$ contains a reaction-diffusion length scale $\sqrt{\DOut/\GammaIn}$, which controls its size.
This behavior is thus reminiscent of Turing patterns~\cite{Menou2023} and other reaction-diffusion systems~\cite{Luo2023}.
In particular, $R_{*,2}$ becomes arbitrarily large for weak reactions (small $\GammaIn$), thus approaching passive droplets.
Taken together, we thus showed how externally-maintained droplets reach a stable stationary size, which also implies that many of such droplets coexist in a large system~\cite{Zwicker2015}.
In this case, the assumption of a large dilute phase can be violated, and the scaling of the droplet size can deviate from the scaling $|\GammaIn|^{1/2}$ predicted by \Eqref{eqn:active_droplet_radius2}~\cite{Christensen1996, Muratov2002}.
A more detailed analysis of externally-maintained droplets becomes possible for a particular case, which we discuss next.

\begin{table}
\begin{center}
\caption{Summary of the behavior of the two classes of chemically active droplets}
\begin{tabularx}{\textwidth}{lll}
\toprule
 & Internally maintained droplet & Externally maintained droplet\\
\midrule
\textbf{Reactions} & Production in droplet, $\GammaIn>0$ &Production outside, $\GammaOut>0$\\
\textbf{Dilute phase}  &Undersaturated, $\eps<0$ &Oversaturated, $\eps>0$
	\\[5pt]
\textbf{Critical radius} &
		$R_\mathrm{c} \approx \left(\frac{3\DOut\epsilon\phiBaseOut}{-\GammaIn}\right)^{\frac12} - \frac{\ell}{\epsilon}$ &
		$R_\mathrm{c} \approx \ell/\epsilon$ \\[15pt]
\textbf{Nucleation} &
	Deterministic by active site~\cite{Zwicker2014} &
	Suppressed by reactions~\cite{Ziethen2023} \\[5pt]
\textbf{Growth} &
	Indefinitely (to system size) &
	Until $R_* \approx \left(\frac{3\DOut\epsilon\phiBaseOut}{-\GammaIn}\right)^{\frac12} - \frac{\ell}{\epsilon}$ \\[15pt]
\textbf{Competition} &
	Accelerated Ostwald Ripening &
	Suppressed ripening~\cite{Zwicker2015} \\
\bottomrule
\end{tabularx}
\label{tab:active_droplets}
\end{center}
\end{table}%

\subsubsection{Fully linearized reactions and electrostatic analogy}
The case of externally-maintained droplets can be discussed for the special case of a constant diffusive mobility $\Lambda(\phi) = \Lambda_0$ and a linear reaction rate law, $s(\phi) = k (\phi_0 - \phi)$, where $\phi_0$ denotes the volume fraction at chemical equilibrium and $k$ denotes the reaction rate; see \figref{fig:active_droplet_reaction_rates}.
If $\phiBaseOut < \phi_0 < \phiBaseIn$, this corresponds to externally-maintained droplets since $\GammaIn = s(\phiBaseOut) > 0$ whereas $\GammaOut < 0$.
Combining this reaction rate law with \Eqref{eqn:dynamics_binary_rate_law} results in the \emph{Cahn-Hilliard-Oono} equation~\cite{Oono1988},
\begin{align}
	\label{eqn:cahn_hilliard_oono}
	\partial_t \phi = \Lambda_0 \nabla^2 \bar\mu + k(\phi_0 - \phi)
	\;,
\end{align}
where $\bar\mu \propto \delta F/\delta \phi$ still denotes the exchange chemical potential of the droplet material, and we consider neutral (Neumann or periodic) boundary conditions.
This equation has the interesting property that the average fraction $\bar\phi = V^{-1} \int \phi \diff V$ satisfies the simple equation $\partial_t \bar\phi =  k(\phi_0 - \bar\phi)$, implying that $\bar\phi$ approaches the average value $\phi_0$ exponentially.

To obtain a concrete form of the equation, we use the free energy functional given by \Eqref{eqn:free_energy_functional} with a simple symmetric free energy density $f_0(\phi) = \frac14 \phi^4 - \frac12\phi^2$, where $\phi$ is now an order parameter which can take any real value.
This leads to the simple equation
\begin{align}
	\partial_t \phi = \nabla^2 \bigl(\phi^3 - \phi - \nabla^2\phi \bigr) + k(\phi_0 - \phi)
\end{align}
where we non-dimensionalized time and space so that the system is described by only the two parameters $k$ and $\phi_0$.
Numerical simulations of this system show that regular patterns with a well-defined length scale emerge~\cite{Glotzer1994a,Glotzer1994}.

To understand the origin of the regular patterns in the Cahn-Hilliard-Oono equation, we next perform a few mathematical transformations on \Eqref{eqn:cahn_hilliard_oono} with the aim to express the reaction term as a correction to the free energy $F$, thus constructing an equilibrium model.
To do this, we first introduce a new field~$\psi$, which satisfies the Helmholtz equation
\begin{align}
	\label{eqn:helmholtz}
	\nabla^2 \psi &= \phi_0 - \phi(\vect r)
	\;,
\end{align}
with appropriate boundary conditions (Neumann or periodic).
Note that \Eqref{eqn:helmholtz} only provides a well-defined solution~$\psi$ if $\int (\phi_0 - \phi) \diff V$ vanishes, but this can be guaranteed by appropriate initial conditions, since the average fraction $\bar\phi$ approaches $\phi_0$ exponentially. 
Using the solution $\psi$, we can turn \Eqref{eqn:cahn_hilliard_oono} into
\begin{align}
	\partial_t \phi = \Lambda_0 \nabla^2\left[\frac{\delta F}{\delta \phi} + \frac{k}{\Lambda_0}\psi(\vect r)\right]
	\;.
\end{align}
We next interpret the square bracket as a re-scaled chemical potential, derived from a re-scaled free energy~$\tilde F$, which reads
\begin{align}
	\label{eqn:free_energy_surrogate_model}
	\tilde F[\phi] &= F[\phi] + \frac{k}{\Lambda_0} \int\bigl[\phi(\vect r) - \phi_0\bigr] \psi(\vect r) \diff \vect r
	\;.
\end{align}
This implies we can write \Eqref{eqn:cahn_hilliard_oono} as
\begin{align}
	\label{eqn:cahn_hilliard_long_ranged}
	\partial_t \phi = \nabla^2 \frac{\delta \tilde F}{\delta \phi}
	\;.
\end{align}
We thus turned the non-equilibrium system described by \Eqref{eqn:cahn_hilliard_oono} into an equilibrium system described by \Eqref{eqn:cahn_hilliard_long_ranged}.
Since both systems are mathematically identically (and thus describe the same dynamics), we can describe the behavior of the Cahn-Hilliard-Oono equation by a passive system with the additional contribution in the free energy~\cite{Christensen1996,Muratov2002, Ziethen2023, Ziethen2024}

\begin{figure}[t]
	\centering	
	\includegraphics[width=0.6\textwidth]{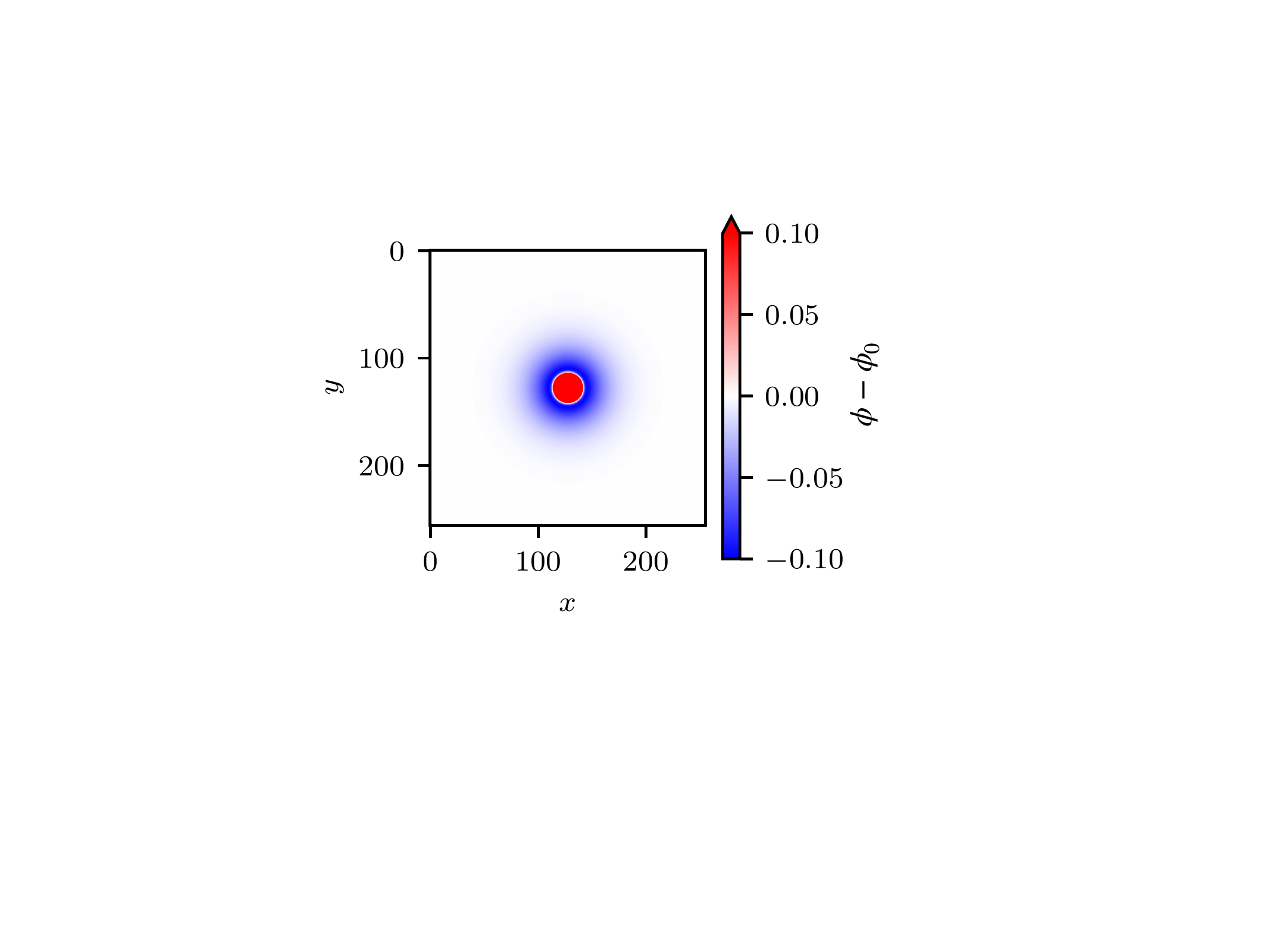}
	\caption{%
	\textbf{Electrostatic analogy of active droplets.}
	The concentration deviation $\phi(\vect r)-\phi_0$ can be interpreted as a charge density, so that the electrostatic interactions have an identical effect to the reactions.
	Picture provided by Noah Ziethen.
	}
	\label{fig:droplet_electrostatics}
\end{figure}

The new term in the free energy must capture the effects of the chemical reactions, and indeed it is proportional to the reaction rate~$k$.
To understand the term in a bit more detail,
we solve \Eqref{eqn:helmholtz} using the Green's functions $G(\vect r, \vect r')$ of the Laplacian,
which satisfies
\begin{align}
	\nabla^2 G(\vect r, \vect r') &= \delta(\vect r - \vect r')
	\;.
\end{align}
The auxiliary field~$\psi$ then reads
\begin{align}
	\psi(\vect r) &= \int \bigl(\phi_0 - \phi(\vect r') \bigr) G(\vect r, \vect r') \diff \vect r'
	\;.
\end{align}
In an infinite 3D system, the Green's function is the Coulomb potential,
\begin{align}
	G(\vect r, \vect r') &=- \frac{1}{4\pi|\vect r - \vect r'|}
	\;.
\end{align}
Consequently, $\psi(\vect r)$ can be interpreted as an electrostatic potential, originating from the deviation $\phi(\vect r) - \phi_0$, which plays the role of the charge density; see \figref{fig:droplet_electrostatics}.
The second term in \Eqref{eqn:free_energy_surrogate_model} is then analogous to the electrostatic energy, completing the picture.
In fact, we can also formally remove the Helmholtz equation~\eqref{eqn:helmholtz} by treating $\psi$ as an independent field,
\begin{align}
	\label{eqn:free_energy_electric_energy}
	\hat F[\phi, \psi] &= F[\phi] +\frac{k}{\Lambda_0}  \int\Bigl[
		 \bigl(\phi(\vect r) - \phi_0\bigr)\psi(\vect r)
		 - |\nabla\psi(\vect r)|^2
	\Bigr] \diff \vect r
	\;.
\end{align}
Minimization of $\hat F$ with respect to $\psi$, i.e., demanding $\delta \hat F/\delta \psi = 0$, yields the Helmholtz equation~\eqref{eqn:helmholtz}, whereas stationary states of $\phi(\vect r)$ follow from $\delta \hat F/\delta \phi = 0$.
The square bracket in \Eqref{eqn:free_energy_electric_energy} denotes the electrostatic energy density.
The advantage of the formulation above is that both fields ($\phi$ and $\psi$) can minimized simultaneously, without solving the Helmholtz equation explicitly.
Taken together, the chemical reactions thus have very similar effects to electrostatics, and lead to long-ranged interactions. 

The electrostatic analogy provides us with an intuitive understanding of the role of reactions in externally-maintained droplets:
Phase separation condenses droplet material in the dense phase, which thus leads to an accumulation of charge.
This behavior alone clarifies that such droplets cannot grow indefinitely since the electrostatic energy would eventually outcompete the surface energy and it is thus favorable to have multiple smaller droplets.
This behavior is known as a Rayleigh instability~\cite{Rayleigh1882} and explains why externally-maintained droplets can split spontaneously~\cite{Zwicker2017,Golestanian2017}.
Moreover, the charged droplets attract a cloud of opposite charge around them, so the entire system is still neutral.
This implies that the interaction of multiple active droplets is screened from each other, similar to interactions in electrolytes.

The fact that the effect of the driven chemical reactions can be captured by long-ranged electrostatic interactions is puzzling.
However, it is important to stress that it is only an analogy.
First, droplet material can additionally carry charges, which affects phase separation, and has nothing to do with electrostatic analogy we presented here~\cite{Duan2023}.
Second, the analogy does not capture the transient dynamics when the initial state does not obey $\bar\phi = \phi_0$.
Third, the analogy only works in the special case of linear chemical reactions and constant mobility; any realistic system will likely exhibit deviations.
Yet, the electrostatic analogy clearly shows that the reaction-diffusion dynamics mediate long-ranged interactions and it is useful for an intuitive understanding of chemically active droplets.

\subsection{Conclusions}

Chemical reactions can modify the phase behavior of mixtures of interacting particles.
In passive systems, adding reactions generally reduces the degrees of freedom, so that fewer coexisting phases are expected.
In contrast to this simple behavior, the external energy input in active systems can sustain diffusive fluxes.
When the flux of droplet material is directed away from droplets, Ostwald ripening is accelerated (in internally-maintained droplets~\cite{Tena-solsona2019}), whereas the opposite flux stabilizes  the droplet size and leads to multiple coexisting droplets (in externally-maintained droplets~\cite{Zwicker2015}); see \tabref{tab:active_droplets}.
In the simplest case, this behavior of externally-maintained droplets is akin to an electrostatic interaction.

\clearpage
\section{Summary}

These lecture notes summarized the basic physics of chemically active droplets, showing how driven reactions can affect standard behavior of droplets.
We necessarily glossed over some details~\cite{Julicher2024}, particularly concerning heat transport in these non-equilibrium systems~\cite{Mabillard2023}, and the details of how fuel is supplied and waste is removed in the driven scenario~\cite{Bauermann2022}.
We here also focused mostly on size-control of chemically active droplets, but the diffusive fluxes caused by the driven reactions also lead to other interesting effects:
Chemically active droplets can form liquid shells~\cite{Bauermann2023,Bergmann2023}, position catalytically-active particles~\cite{Zwicker2018b}, and control droplet nucleation~\cite{Ziethen2023, Ziethen2024}.
Such droplets also respond to external gradients~\cite{JambonPuillet2023}, and they can even exhibit spontaneous symmetry-breaking, leading to self-propulsion~\cite{Demarchi2023}.
These effects demonstrate how chemical reactions affect phase separation, but the tight coupling between the two physical processes implies that phase separation can also have tremendous effects on chemical reactions:
It is now clear that concentrating reactants and products in droplets can have profound effect on the yield of reactions~\cite{Laha2024} and the assembly of oligomers or fibers~\cite{Bartolucci2023}.
Taken together, all these effects demonstrate that driven reactions provide a flexible mechanism for controlling droplets.

Beyond chemical reactions there exist various other control mechanisms that influence droplets.
For instance, droplets can interact with solid structures, leading to wetting phenomena.
The relative surface tensions determine the contact angle of the droplet with the solid surface according to the Young–Dupré equation, whereas the droplet shape is still governed by the energy minimization demanding a constant mean curvature.
Chemical activity can deform such sessile droplets~\cite{Ziethen2024}, and surfaces can also bind reactants and localize reactions, leading to more complex phenomena~\cite{Zhao2024,Zhao2021}.
Droplets can also be affected by controlling external parameters, like temperature or chemical potentials of reservoirs, which in turn can affect chemical reactions~\cite{Haugerud2023}.
Droplets also respond to heterogeneities, like external gradients in concentration~\cite{Weber2017} or material properties~\cite{Boddeker2023a, Vidal2020, Rosowski2019}.
The local elastic environment of droplets can also have a profound effect on droplet formation, either because it controls cavitation~\cite{Vidal2021}, or because it constrains droplets to softer regions~\cite{Qiang2023, Fernandez-Rico2023, Rosowski2019}.
These selected examples indicate that chemically active droplets show a rich interplay with other physical processes that we are only beginning to understand.
More research in this direction is necessary to understand the intricate processes that control biomolecular condensates in cells and to apply similar principles in engineering to design smart materials.

\clearpage
\begin{appendix}
\section{Concrete expressions for binary mixtures}
\label{app:binary_mixtures}
This appendix collects some results on the thermodynamics of binary mixtures.
For incompressible systems, the system's state is specified by the volume fraction~$\phi$ of one of the components.
The free energy density can then be written as
\begin{equation}
	\label{eqn:free_energy_binary}
	f(\phi) = 
		\frac{\kBT}{\nu} \Bigl[
			\phi \ln(\phi)
			+ (1-\phi) \ln(1 - \phi)
			+ h(\phi)
		\Bigr]
	\;,
\end{equation}
where the first two terms capture the translational entropies of the two components, whereas the enthalpic contribution $h(\phi)$ captures all other contributions to the energy.
Using a free energy $F[\phi] = \int [f(\phi) + \frac\kappa2|\nabla \phi|^2]\diff V$, the exchange chemical potential reads
\begin{align}
	\label{eqn:chemical_potential_binary}
	\bar\mu(\phi) = \kBT\Bigl[
		\ln(\phi) - \ln(1 - \phi) + h'(\phi)
	\Bigr]
	- \kappa \nabla^2 \phi
	\;,
\end{align}
and the Osmotic pressure is
\begin{align}
	\label{eqn:pressure_binary}
	\bar\Pi 
	= \frac{\kBT}{\nu}\Bigl[		 
		\phi h'(\phi) - h(\phi)
		-\ln(1 - \phi)
	\Bigr]
	\;.
\end{align}
\paragraph{The Flory-Huggins free energy}
provides a  concrete choice for $h(\phi)$,
\begin{align}
	h(\phi) &= 
		 w \phi
		+ \chi \phi(1-\phi)
& \Rightarrow &&
	h'(\phi) &= 
		 w 
		+ \chi (1-2\phi)
	\;,
\end{align}
where $w$ is the internal energy difference between the two particle types (which does not affect phase separation), whereas $\chi$ captures their interaction (to lowest order).
Homogeneous states (without reactions) are then stable if $\chi$ is smaller than
\begin{equation}
	\label{eqn:chi_spinodal}
	\chi_\mathrm{spin}(\phi) 
	= \frac12\left(\frac{1}{\phi}  + \frac{1}{1- \phi}\right)
	\;,
\end{equation}
which marks the spinodal line.
The minimum at $\chi_*=2$ and $\phi_*=\frac12$ denotes the critical point.
A phase separated state is stable if $\chi$ is larger than
\begin{equation}
	\label{eqn:chi_binodal_n1}
	\chi_\mathrm{bin}(\phi) = \frac{\ln (1-\phi )-\ln (\phi )}{1-2 \phi }
	\;,
\end{equation}
which thus corresponds to the binodal line.
The coexisting volume fractions in the two phases can be estimated using a fixed-point iteration~\cite{Qian2022},
\begin{align}
	\label{eqn:phi_binodal_approx}
	\phiBase_\mathrm{in/out} &\approx \frac{1}{1 + \exp\left[-\chi \tanh\left(\pm \chi \sqrt{\frac{3\chi - 6}{8}}\right)
	\right]}
	\;.
\end{align}

\end{appendix}

\clearpage
\small
\bibliographystyle{unsrtnat}
\bibliography{bibdesk}

\end{document}